\documentclass[11pt]{article}
\newcommand{\Comment}[1]{{}}
\usepackage[textwidth = 430 pt, textheight = 630 pt]{geometry}
\usepackage{amsmath,euscript,amssymb,amsfonts,graphicx,bbm}

\usepackage{color}
\usepackage{bbold}
\definecolor{MyDarkBlue}{rgb}{0.15,0.15,0.45}
\usepackage[linktocpage=true]{hyperref}
\hypersetup{
colorlinks=true,
citecolor=blue,
linkcolor=blue,
urlcolor=blue,
pdfauthor={Andrew~B.~ Royston},
pdftitle={Monopole-defect simulations},
pdfsubject={hep-th}
}

\usepackage[numbers,sort&compress]{natbib}
\usepackage{hypernat}
\usepackage{slashed}

\DeclareMathAlphabet{\mathpzc}{OT1}{pzc}{m}{it}
\def\cg{\mathpzc{g}}
\def\pp{\mathpzc{p}}
\def\qq{\mathpzc{q}}

\newcommand\ignore[1]{}
\def\half{\frac{1}{2}}

\newcommand{\sgn}{\operatorname{sgn}}

\newcommand{\Tr}{\operatorname{Tr}}

\newcommand{\csch}{\operatorname{csch}}

\newcommand{\pd}{\partial}
\newcommand{\ie}{{\it i.e.~}} 
\newcommand{\eg}{{\it e.g.~}}
\newcommand{\etc}{{\it etc.~}}

\def\MM{\mathcal{M}}
\def\PP{\mathcal{P}}
\def\YY{\mathcal{Y}}
\def\RR{\mathcal{R}}

\def\gym{{g_{\textrm{ym}}}}
\def\gymsq{{g_{\textrm{ym}}^2}}
\def\thetaym{{\theta_{\textrm{ym}}}}
\def\tthetaym{{\tilde{\theta}_{\textrm{ym}}}}

\begin{document}

 
   \vspace*{3truecm}

\centerline{\LARGE \bf {\sc Simulating Magnetic Monopole-Defect Dynamics}} 
\vspace{1.5truecm}
\centerline{ {\large {\bf{\sc Gannon E.~Lenhart,}${}^{\,a,}$}}\footnote{E-mail address:
    \href{mailto:Gannon Lenhart
      <gkl5103@psu.edu>}{\tt
      gkl5103@psu.edu}. 
      Current Affiliation:  Department of Aerospace Engineering, The Pennsylvania State University, University Park, PA 16802, USA} {\large {\bf{\sc
        Andrew~B.~Royston}${}^{\,a,}$}}\footnote{E-mail address:
    \href{mailto:Andy Royston <abr84@psu.eduu>}{\tt
      abr84@psu.edu} } {and} {\large {\bf{\sc Keaton E.~Wright}${}^{\,a,}$}}\footnote{E-mail address:
    \href{mailto:Keaton Wright
      <kew5583@psu.edu>}{\tt
      kew5583@psu.edu}.
      Current Affiliation:  Department of Physics, The Pennsylvania State University, University Park, PA 16802, USA} }

\vspace{1cm}
\centerline{${}^a${\it Department of Physics, Penn State Fayette, The Eberly Campus}}
\centerline{{\it 2201 University Drive, Lemont Furnace, PA 15456, USA}}

\vspace{1.5truecm}

\thispagestyle{empty}

\centerline{\sc Abstract}
\vspace{0.4truecm}
\begin{center}
  \begin{minipage}[c]{380pt}{We present simulations of one magnetic monopole interacting with multiple magnetic singularities.  Three-dimensional plots of the energy density are constructed from explicit solutions to the Bogomolny equation obtained by Blair, Cherkis, and Durcan.  Animations follow trajectories derived from collective coordinate mechanics on the multi-centered Taub--NUT monopole moduli space.  We supplement our numerical results with a complete analytic treatment of the single-defect case.}
\end{minipage}
\end{center}

\vspace{.4truecm}

\noindent

\vspace{.5cm}

\setcounter{page}{0}

\newpage

\renewcommand{\thefootnote}{\arabic{footnote}}
\setcounter{footnote}{0}

\linespread{1.1}
\parskip 7pt

{}~
{}~

\makeatletter
\@addtoreset{equation}{section}
\makeatother
\renewcommand{\theequation}{\thesection.\arabic{equation}}

\tableofcontents

\section{Introduction and Summary}

A standard course in introductory physics can't help but suggest a false dichotomy, with particle mechanics---or more generally the mechanics of rigid bodies---on the one side, and field theory on the other side.  One learns that charged particles create fields and fields apply forces to charged particles, but one is not presented with a complete description of a coupled particle-field system until later.  Furthermore, the fields are typically singular in the vicinity of the particles sourcing them, so that the dichotomy is only truly resolved through quantum field theory.  This is the situation, at least, for electromagnetism and its charged particles.

Solitons provide a fascinating alternative if one's goal is to see how particle dynamics can emerge from, and be completely embedded in, the framework of classical field theory in a nonsingular way.  Static solitons are represented by self-supporting localized field configurations and occur in theories that admit a topological charge.\footnote{In theories on flat, infinitely-extended Euclidean space, values of the topological charge label different classes of boundary conditions at spatial infinity consistent with finiteness of the energy.  Two field configurations with asymptotic boundary conditions in the same class can be smoothly deformed into each other through a sequence of finite energy configurations; two configurations from different classes cannot.  A soliton solution is an energy-minimizing field configuration within a (nontrivial) topological class.}  Solitons exist in a variety of theories; classic examples include kinks (or domain walls) for theories in one spatial dimension, vortices for theories in two dimensions, and magnetic monopoles for theories in three dimensions.  See \cite{MR2068924} for a modern review.

The existence of solitons typically relies on nonlinearity in the field equations, and one cannot linearly superpose two one-soliton solutions to construct a two-soliton solution.  Nevertheless, soliton solutions do come in smooth families with a number of parameters, or \emph{moduli}, parameterizing the family.  Focusing on the case of magnetic monopoles in Yang--Mills--Higgs theory \cite{tHooft:1974kcl,Polyakov:1974ek}, solutions representing a single monopole occur in a four-dimensional family.  Three moduli represent the position of the monopole in $\mathbbm{R}^3$ while the fourth parameter is a circle coordinate, with momentum along this circle corresponding to electric charge \cite{Julia:1975ff}.  This four-dimensional space of solutions is referred to as the one-monopole moduli space.

In this paper we work in the context of a special type of Yang--Mills--Higgs theory---namely, one which has (four-dimensional $\mathcal{N} = 2$ extended) supersymmetry \cite{Ferrara:1974pu,Fayet:1975yi,Grimm:1977xp,Witten:1978mh,Seiberg:1994rs}.  In supersymmetric Yang--Mills--Higgs theory, the net force vanishes between two stationary monopoles and hence there exist static $n$-monopole solutions coming in smooth $4n$-dimensional families for any positive integer $n$.\footnote{This no-force condition also arises in ordinary Yang--Mills--Higgs theory when the Higgs self-coupling is tuned to zero, a limit first considered by Bogomolny \cite{Bogomolny:1975de}, and Prasad and Sommerfield \cite{Prasad:1975kr}; hence the common terminology ``BPS monopole.''}  These monopole moduli spaces have been intensely studied by mathematicians and physicists alike since their invention in the late 1970's.  They inherit a natural Riemannian metric induced from the energy functional of the parent Yang--Mills--Higgs theory that carries rather special geometric structures.  Specifically, these moduli spaces are hyperk\"ahler manifolds admitting a number of isometries; see \cite{Atiyah:1988jp} for details.  In \cite{Atiyah:1988jp}, Atiyah and Hitchin explotied these special structures to pin down the metric on the two-monopole moduli space, despite the fact that the full family of two-monopole solutions was not known at the time.\footnote{Although explicit solutions were not known, one could still prove they exist \cite{MR614447}.  The fully explicit analytic solution for the general two-monopole configuration was obtained quite recently in \cite{Braden:2019ijz}, building on many earlier efforts and partial results.}

Furthermore, the collective coordinate paradigm of Manton \cite{Manton:1981mp} reaches its full brilliance in the context of multi-monopole moduli spaces.  Collective coordinates are the ultimate example of the physicist's ball-rolling-on-a-hill.  In this analogy, the ball is the multi-monopole configuration, the terrain of peaks and valleys is the infinite-dimensional space of field configurations, and monopole moduli space is a minimum-energy valley where the ball can roll without change in kinetic energy.  In fact the analogy is rigorous.  It can be proven that time-dependent solutions to the full field equations are well-approximated by allowing the moduli to become time-dependent -- \ie promoted to collective coordinates -- so that they trace out specific trajectories in moduli space \cite{Stuart1994}.  

The trajectories in moduli space are determined by a specific form of Newton's Laws, and hence the emergence of particle mechanics from field theory.  In ordinary Yang--Mills--Higgs theory in the Bogomolny--Prasad--Sommerfield (BPS) limit, Newton's Laws imply that the trajectories are geodesics \cite{Manton:1981mp}.  However, in the supersymmetric extension considered here, there is an additional force due to a secondary Higgs field that modifies the trajectories in a way compatible with the special structure of the moduli space \cite{Gauntlett:1999vc,Gauntlett:2000ks}.\footnote{Specifically, the force follows from a potential energy function on moduli space that is the norm-squared of a tri-holomorphic Killing vector field.  The latter is a vector field that generates isometries preserving the hyperk\"ahler structure of the moduli space.}

Motion on the two-monopole moduli space was studied in \cite{Atiyah:1988jp}, where it was shown to predict rather beautiful and dramatic phenomena for the scattering of two monopoles in real space.  For example, in a head-on collision, the two monopoles---represented by spherical blobs of energy when they are far apart---deform as they get close to each other.  The monopoles' individual identities disappear momentarily as they overlap and form an axially symmetric ring of energy.  The spherical blobs then re-emerge from the collision region traveling away from each other on a line rotated by 90 degress from the line of incidence.  A simulation of the collision \cite{Merlinetal}, constructed in the late 1980's on an IBM supercomputer, can still be found on YouTube.   Very recently, an interactive applet has been constructed for the two monopole solution based on the new analytic results of \cite{Braden:2019ijz}.  See the final appendix of Reference \cite{Braden:2019ijz} for discussion and links. 

Fascinating $n$-monopole collisions with $n > 2$ have been studied for special initial conditions such that the configuration maintains some specific symmetry throughout the evolution; see \emph{e.g.}~\cite{Houghton:1995bs,Houghton:1995uj}.  The reason for this symmetry restriction is that the full moduli space geometry, required to simulate collisions with generic initial conditions, is not known for $n > 2$.  

Our goal in this work is to explore and simulate multi-monopole interactions in a different limit---namely, when all but one of the monopoles are infinitely heavy and immobile, while the remaining one can move in accordance with the appropriate moduli space force law.  The heavy monopoles can be placed at arbitrary fixed positions in three dimensional space and are modeled as magnetic singularities known as (supersymmetric) 't Hooft defects. \cite{'tHooft:1977hy,Kapustin:2005py}.  They can indeed be viewed as infinite-mass limits of ordinary monopoles in a precise sense described in \cite{MRVdimP2,Brennan:2018ura}.  This allows us to utilize a relatively recent set of analytic solutions obtained by Blair, Cherkis, and Durcan (BCD) \cite{Cherkis:2007jm,Cherkis:2007qa,Blair:2010kz,Blair:2010vh}, describing one ordinary mobile monopole in the presence of any number of fixed 't Hooft defects.  We compute the energy density and construct three-dimensional plots, using Mathematica, for the smooth monopole in arbitrary position relative to the defects.  Like the authors of reference \cite{Braden:2019ijz}, we plot several level sets of the energy density with varying opacity, so as to allow one to see inside the monopole configuration.  See Figure \ref{fig:EnergyPlots} below.

The moduli space metric for one monopole in the presence of $k$ minimally charged singularities is also known and given by the $k$-centered Taub--NUT manifold \cite{Cherkis:1997aa}.  This is a four-dimensional manifold constructed over an $\mathbbm{R}^3$ base.  Over each point on the base is a fiber that is generically a circle.  However the size of the circle varies, and the circle shrinks to a point over the $k$ ``nut'' points of the base, whose positions are specified by $k$ fixed three-dimensional vectors.  The moduli parameterizing the three-dimensional base correspond to the position of the smooth monopole in physical three-space, and the nut points represent the positions of the singularities.  Momentum along the circle fiber represents electric charge carried by the monopole.  There is also a potential energy function on the moduli space due to the secondary Higgs field in the supersymmetric model that we consider.

We numerically integrate Newton's Laws with Mathematica to determine the trajectory on moduli space for any given initial position and velocity.  We then use Manton's collective coordinate ansatz and the explicit BCD solutions to create the resulting simulations of the monopole interacting with the defects in real space.  There exist both bound orbits (generically non-repeating, but closed for special initial conditions) and unbound trajectories.  When the smooth monopole passes near the defect we observe significant but transitory deformations of both the monopole and defect shape.  Complete, momentary screening of the defect by the monopole is also observed when the defect carries the same charge as the monopole.

When only a single defect is present, the equations of motion on moduli space can be integrated analytically for generic initial conditions.  We carry out this analysis as well, since the analytic results offer valuable insight into the more complex scenarios with multiple defects.  This system is mathematically equivalent to several related systems that have been studied over the years, starting with work of Zwanziger \cite{Zwanziger:1969by,Lee:2000rp,Jante:2015xra}.  In these references it was found that trajectories are conic sections.  However, the plane of motion does not contain the defect when the monopole carries electric charge.  We review and extend some of these results to include the period of bound orbits and a new and elementary approach to the classical scattering problem.

The structure of the paper is as follows.  In section \ref{sec:gaugetheory} we review the theoretical background for the emergence of monopole-defect solutions and dynamics in supersymmetric Yang--Mills--Higgs theory.  Then in section \ref{sec:BCD} we apply these ideas to the explicit BCD solutions and illustrate them with three-dimensional energy density plots and simulations based on moduli-space dynamics.  Section \ref{sec:Analytic} contains our analysis of the single-defect case.  We conclude in section \ref{sec:Conclusions} with a brief summary and description of future directions.

Several simulations are highlighted in subsections \ref{ssec:sim}, \ref{ssec:timedep}, and \ref{ssec:scattering}.  These simulations, and the Mathematica code required to create such simulations, are included with the arXiv submission of this manuscript as ancillary files.  High resolution simulations can be created in a few hours to a couple days on current commercial laptops, depending on the number of defects included.  Low resolution simulations can be made in a matter of minutes.

\section{Monopole-defect Dynamics in Gauge Theory}\label{sec:gaugetheory}

\subsection{Supersymmetric Yang--Mills--Higgs with 't Hooft Defects}

We study a field theory on Minkowski space, $\mathbbm{R}^{1,3}$ with coordinates $x^\mu = (t,\vec{r})$, consisting of a non-abelian gauge field $A_\mu = (A_0,A_i)$, $i=1,2,3$, two adjoint-valued scalars, $X,Y$, and a pair of adjoint-valued Weyl fermions.  Although the fermions are crucial for supersymmetry, they will not be utilized in the following and so we suppress their contribution to the action and Hamiltonian below.  We work with the simplest nonabelian gauge group, $G = \mathrm{SO}(3)$.  The three generators of the Lie algebra, $\mathfrak{g} = \mathfrak{so}(3)$, are denoted $T^a$, $a = 1,2,3$.  We use anti-Hermitian generators satisfying $[T^a,T^b] = \epsilon^{ab}_{\phantom{ab}c} T^c$ and normalized such that $\textrm{Tr}(T^a T^b) = \half \delta^{ab}$.\footnote{Here ``Tr'' denotes a positive-definite Killing form on the Lie algebra.  We work in the minimal two-dimensional representation for $\mathfrak{so}(3)$ where $T^a = -\frac{i}{2} \sigma^a$ with $\sigma^a$ the Pauli matrices.  Then Tr is the negative of the ordinary matrix trace.}  Each field can be expanded in this basis; $A_i = A_{ia} T^a$, $X = X_a T^a$, \etc  The covariant derivative and non-abelian field strength tensor are 
\begin{equation}
D_\mu X = \pd_\mu X + [A_\mu, X]~, \qquad F_{\mu\nu} = \pd_{\mu} A_\nu - \pd_\nu A_\mu + [A_\mu, A_\nu]~.
\end{equation}
The magnetic field is $B_i = \half \epsilon_{ijk} F^{jk}$, and we work in mostly plus conventions such that the electric field is $E_i = F_{i0}$.

Gauge transformations act on the fields according to $(A_\mu, X, Y) \mapsto (A_{\mu}', X', Y')$ with
\begin{equation}
A_{\mu}' = \cg (A_\mu + \pd_\mu) \cg^{-1} ~, \qquad X' = \cg X \cg^{-1} ~, \qquad Y' = \cg Y \cg^{-1}~.
\end{equation}
where $\cg(x) \in \mathrm{SO}(3)$.  Taking $\cg = \exp(\epsilon_a T^a)$, the infinitesimal form of these transformations is
\begin{equation}
\delta A_\mu = - D_\mu \epsilon~, \qquad \delta X = [\epsilon, X]~,\qquad \delta Y = [\epsilon, Y]~.
\end{equation}
Two field configurations related by a local gauge transformation, \ie a gauge transformation with $\cg \to \mathbbm{1}$ (or $\epsilon \to 0$) as $\vec{r} \in \mathbbm{R}^3$ is sent to infinity, are physically equivalent.  In contrast, global gauge transformations---those that do not approach the identity at spatial infinity---generate symmetries that can be used to simplify asymptotic boundary conditions and generate conserved Noether charges.

We probe this theory with magnetic defects known as 't Hooft defects \cite{'tHooft:1977hy}, which can be thought of as magnetic duals to the Wilson lines of external electrically charged particles.  't Hooft defects are a type of disorder operator, in that they are defined not in terms of the local fields in the theory but rather in terms of singular boundary conditions on the fields.  This construction was extended to the supersymmetric context in \cite{Kapustin:2005py}.  Such supersymmetric ``line defects'' have played a central role in many of the new theoretical developments of the past decade for supersymmetric Yang--Mills--Higgs theory, beginning in large part with the work of Gaiotto, Moore, and Neitzke \cite{Gaiotto:2010be}.  Reference \cite{Moore:2015szp} analyzed the semiclassical description of magnetic defects in supersymmetric gauge theory and the connection to singular monopoles.  We refer the reader to \cite{Moore:2015szp} for details and further discussion of the results reviewed here.

A supersymmetric 't Hooft defect placed at position $\vec{\nu}_J \in \mathbbm{R}^3$ is specified by a charge $P_J$ and defined by imposing the singular boundary conditions
\begin{equation}\label{tHooftdefect}
\vec{B} = \frac{P_J}{r_{J}^2} \hat{r}_J + \cdots ~, \qquad X = -\frac{P_J}{2r_J} + \cdots ~, \qquad \textrm{as} \quad r_J \equiv | \vec{r} - \vec{\nu}_J| \to 0~,
\end{equation}
where the ellipses represent subleading terms.  By making local gauge transformations, the charge $P_J$ can be taken to be a constant element of the Lie algebra, valued in a Cartan subalgebra.  For $\mathfrak{so}(3)$ we take this Cartan subalgebra to be generated by $T^3$, so that $P_J$ is specified by a single integer: 
\begin{equation}\label{tHooftcharge}
P_J = p_J T^3~, \qquad |p_J| =1,2,3,\ldots~.
\end{equation}
Hence 't Hooft defects may be thought of as Dirac monopoles, where the 't Hooft charge, $P$, specifies an embedding of the magnetic charge into the non-abelian gauge group.  Dirac quantization restricts $p_J$ to be an integer.  Since gauge transformations can be used to send $P_J \to -P_J$, it is only $|p_J|$ that is physical.  

The action for supersymmetric Yang--Mills--Higgs theory in the presence of some number of 't Hooft defects placed at positions $\{ \vec{\nu}_J \}$ is a sum of two types of terms.  The first set of terms is referred to as the ``vanilla'' action in \cite{Moore:2015szp}, and comprises those terms that would ordinarily be present for the theory without defects.  The vanilla action depends on two parameters, the Yang--Mills coupling $g_{\rm ym}$ and the theta angle $\theta_{\rm ym}$.  The second set of terms are boundary terms supported on infinitesimal two-spheres, $S_{J}^2$, surrounding the defects.  These terms are required for consistency of the variational principle and preservation of supersymmetry.  Letting $\varepsilon_{J}$ denote the radius of $S_{J}^2$ and $\Omega_J$ the solid angle, one has
\begin{align}
S =&~ S_{\rm van} + S_{\rm def} \equiv \int dt L~, \qquad \textrm{where} \label{Stot} \\
S_{\rm van} =&~ - \frac{1}{\gymsq} \int d^4 x \Tr \left\{ \half F_{\mu\nu} F^{\mu\nu} + D_\mu X D^\mu X + D_{\mu} Y D^{\mu} Y + [X,Y]^2 \right\} + \textrm{fermions}~ + \cr
&~ - \frac{\thetaym}{32\pi^2} \int d^4 x \epsilon^{\mu\nu\rho\sigma} \Tr(F_{\mu\nu} F_{\rho\sigma}) ~, \label{Svan} \raisetag{24pt} \\
S_{\rm def} =&~ -\int dt \sum_{J} V_{\rm def}^{(J)}~, \label{Sdef}
\end{align}
with
\begin{equation}\label{Vdef}
V_{\rm def}^{(J)} = \frac{2}{\gymsq} \int_{S_{J}^2} d^2 \Omega_J \varepsilon_{J}^2 \hat{r}_{J}^i \Tr ( X B_i + Y E_i) ~.
\end{equation}

The integrals over space in $S_{\rm van}$ should be taken to exclude the infinitesimal balls bounded by the $S_{J}^2$'s surrounding the defect insertions.  The $\theta_{\rm ym}$ term can be written as a total derivative and ordinarily does not contribute to the dynamics, but in the presence of defects the additional boundaries enable this term to become dynamical, as we will see below.  When $\theta_{\rm ym} \neq 0$, the 't Hooft defect is a source for the electric field and $Y$ scalar as well, with these fields behaving as \cite{Moore:2015szp}
\begin{equation}
\vec{E} = - \tthetaym \cdot \frac{P_J}{r_{J}^2} \hat{r}_J + \cdots ~, \qquad Y = \tthetaym \cdot \frac{P_J}{2r_J} + \cdots ~, \quad \textrm{as} \quad r_J \equiv | \vec{r} - \vec{\nu}_J| \to 0~,
\end{equation}
in the vicinity of the defect, where
\begin{equation}
\tthetaym = \frac{\gymsq \thetaym}{8\pi^2}
\end{equation}
sets the relative scale of the $F_{\mu\nu} F^{\mu\nu}$ and $\half \epsilon^{\mu\nu\rho\sigma} F_{\mu\nu} F_{\rho\sigma}$ terms in \eqref{Svan}.  This can be viewed as a manifestation of the Witten effect \cite{Witten:1979ey} for line defects.

\subsection{Magnetic Monopoles and the BPS Bound}

Setting the fermions to zero, the Hamiltonian, or energy functional, associated with the action \eqref{Stot} is
\begin{align}\label{ftHam}
H =&~ \frac{1}{\gymsq} \int_{\mathcal{U}} d^3 x \mathcal{E}+ \sum_J V_{\rm def}^{(J)} ~,
\end{align}
with local energy density
\begin{equation}\label{endenfull}
\mathcal{E} = \Tr \left\{ E_i E^i + B_i B^i + (D_0 X)^2 + D_i X D^i X + (D_0 Y)^2 + D_i Y D^i Y + [X,Y]^2 \right\} ~.
\end{equation}
Here $\mathcal{U}$ is $\mathbbm{R}^3$ with the infinitesimal balls around the defects removed.  This result for the Hamiltonian holds provided that the Gauss Law constraint (or $A_0$ equation of motion),
\begin{equation}\label{Gauss}
D^i E_i - [X, D_0 X] - [Y, D_0 Y] = 0~,
\end{equation}
is imposed.  This constraint is preserved by the time evolution as a consequence of gauge invariance.

The conditions for energy-minimizing field configurations are exposed by rewriting the Hamiltonian \eqref{ftHam} as a sum of squares.   Using integration by parts, cyclicity of the trace, the Gauss Law constraint \eqref{Gauss}, and the Bianchi identity $\epsilon^{\mu\nu\rho\sigma} D_\nu F_{\rho\sigma} = 0$, one finds that \eqref{ftHam} can be written as\footnote{This type of manipulation is often referred to as a `Bogomolny trick' due to its appearance in \cite{Bogomolny:1975de}, though the same type of manipulation was used earlier in \cite{Belavin:1975fg}.}
\begin{align}\label{ftHam2}
H =&~ \frac{1}{\gymsq} \int_{\mathcal{U}} d^3 x \Tr \bigg\{ \left( B_i - D_i X \right)^2 + \left( E_i - D_i Y \right)^2 + (D_0 Y)^2 + \left( D_0 X + [X,Y] \right)^2 \bigg\} + M~,
\end{align}
where $M$ is a boundary term receiving contributions from the two-sphere at spatial infinity:
\begin{equation}\label{BPSmass1}
M = \frac{2}{\gymsq} \int_{S_{\infty}^2} d^2 \Omega \lim_{r \to \infty} \left\{ r^2 \hat{r}^i \Tr( X B_i + Y E_i) \right\} ~.
\end{equation}
Here, the $V_{\rm def}^{(J)}$ terms in \eqref{ftHam} cancel boundary terms generated from integration by parts on the infinitesimal two-spheres surrounding the defects, so that only the two-sphere at infinity contributes to $M$.

Asymptotic boundary conditions can be chosen to ensure finiteness of the energy.   We require the magnetic and electric fields to fall off like $O(1/r^2)$ while the Higgs fields must become covariantly constant, mutually commuting, and must also commute with the $O(1/r^2)$ terms of the electric and magnetic fields.  By a suitable gauge transformation we can assume
\begin{align}\label{asymptotics}
\begin{array}{l l}  \vec{B} = \dfrac{\gamma_{\rm m}}{2 r^2} \hat{r} + \cdots ~, & X = X_\infty - \dfrac{\gamma_{\rm m}}{2r} + \cdots \\[2ex]
\vec{E} = \dfrac{ \gymsq \gamma_{\rm e}^{\rm phys}}{8\pi r^2} \hat{r} + \cdots ~,  & Y = Y_\infty - \dfrac{ \gymsq \gamma_{\rm e}^{\rm phys}}{8\pi r} + \cdots ~, \end{array} \quad r = |\vec{r}| \to \infty~,
\end{align}
where $\gamma_{\rm m}, \gamma_{\rm e}^{\rm phys}, X_\infty, Y_\infty$ are all constants, valued in the same Cartan subalgebra.  The boundary term \eqref{BPSmass1} can be evaluated in terms of the asymptotic data:
\begin{equation}\label{BPSmass}
M = \frac{4\pi}{\gymsq} \Tr (X_\infty \gamma_{\rm m}) + \Tr (Y_\infty \gamma_{\rm e}^{\rm phys}) ~.
\end{equation}

Nonzero values of $X_{\infty}$ or $Y_{\infty}$ indicate spontaneous symmetry breaking for the theory, in which the vacuum breaks the gauge symmetry to $\mathrm{U}(1) \subset \mathrm{SO}(3)$.  The $\mathrm{U}(1)$ fields are those along the Cartan direction and can be associated to ordinary electromagnetism, while the remaining gauge fields---the $W$-boson and its conjugate---receive masses from $X_\infty,Y_\infty$.  Specifically, if we set
\begin{equation}
m_X := \sqrt{2 \Tr(X_{\infty}^2)}~, \qquad m_Y := \sqrt{2 \Tr(Y_{\infty}^2)}~,
\end{equation}
then the mass-squared of the $W$-boson is $m^2 = m_{X}^2 + m_{Y}^2$.  Meanwhile, $\gamma_{\rm e}^{\rm phys}$ is the (physical) electric charge in the system as measured by the flux of the electric field through the two-sphere at infinity.  The notation $\gamma_{\rm e}$ is reserved for the charge of the Noether current associated with global gauge transformations that preserve the vacuum.  The two are different when $\thetaym \neq 0$:
\begin{equation}\label{echargeshift}
\gamma_{\rm e}^{\rm phys} = - \left( \gamma_{\rm e} + \frac{\thetaym}{2\pi} \gamma_{\rm m} \right)~;
\end{equation}
see \cite{Moore:2015szp} for details.

The magnetic charge, $\gamma_{\rm m}$, may include contributions from ordinary monopoles in addition to the magnetic singularities.  The presence of such monopoles requires $m_X \neq 0$.  The allowed values of $\gamma_{\rm m}$ are constrained by both topology and dynamics.  The condition for any simple Lie group and set of 't Hooft defects was determined in \cite{Moore:2014jfa}, building on earlier works \cite{Weinberg:1979ma,Kronheimer,MR1624279,Kapustin:2006pk}, and here we state the result for $\mathrm{SO}(3)$.  Letting $T^3$ be the generator of the Cartan subalgebra defined by $X_\infty$, such that
\begin{equation}\label{Higgsvevs}
X_\infty = m_X T^3~, \qquad Y_\infty = m_Y T^3~,
\end{equation}
we have
\begin{equation}\label{magcharge}
\gamma_{\rm m} = \left(2 n_{\rm m} - \sum_J |p_J| \right) T^3~, \qquad n_{\rm m} \in \{ 0, 1, 2,\ldots \}~,
\end{equation}
where the $p_J$ determine the 't Hooft charges, \eqref{tHooftcharge}.  The non-negative integer $n_{\rm m}$ is the number of ordinary monopoles present in the system.

The expression \eqref{ftHam2} implies the lower bound on the energy functional,
\begin{equation}
H \geq M~,
\end{equation}
for a given set of asymptotic boundary conditions.  The bound is saturated when all of the squares in \eqref{ftHam2} vanish, leading to the Bogomolny--Prasad--Sommerfield (BPS) equations
\begin{equation}\label{BPSeqns}
B_i - D_i X = 0~, \quad E_i - D_i Y = 0~, \quad D_0 Y = 0~, \quad D_0 X + [X,Y] = 0~.
\end{equation}
A solution to these equations and the Gauss Law constraint will automatically solve the full equations of motion.  

A convenient gauge choice for studying solutions to \eqref{BPSeqns} is $A_0 = Y$, in which case the last three equations imply $A_i,X,Y$ are time-independent.  This leaves only the first equation, which we recognize as Bogomolny's equation for magnetic monopoles \cite{Bogomolny:1975de}, and the Gauss Law constraint.  The constraint can be rewritten, using the latter three of \eqref{BPSeqns}, as a linear equation for $Y$ in a background $(A_i,X)$ that solves the Bogomolny equation:
\begin{equation}\label{BPSeqns2}
B_i = D_i X~, \qquad D^i D_i Y + [X,[X,Y]] = 0~.
\end{equation}
Thus we see that the Bogomolny equation arises as an energy minimizing condition.  One is interested in solutions to these equations modulo gauge transformations that preserve the condition $A_0 = Y$.  These are simply the time-independent gauge transformations.

Finding solutions to the Bogomolny equation, with singularities of the form \eqref{tHooftdefect} and asymptotics of the form \eqref{asymptotics}, is a well-studied problem going back to \cite{Kronheimer}.  While explicit solutions are rare, a great deal is known about general properties of the space of gauge-inequivalent solutions.  This space is referred to as the \emph{moduli space of singular monopoles}.   For $G = \mathrm{SO}(3)$, with 't Hooft defects carrying charges $P_J$, \eqref{tHooftcharge}, and an asymptotic magnetic charge given by \eqref{magcharge}, the moduli space of singular monopoles will be denoted $\MM(n_{\rm m}, m_X; \{ p_J, \vec{\nu}_J\} )$.  We will often use the shorthand $\MM$ when the context is clear.\footnote{The notation $\overline{\underline{\mathcal{M}}}$ was used for singular monopole moduli space in \cite{Moore:2015szp} to distinguish it from the ordinary monopole moduli spaces that occur in the absence of 't Hooft defects.  We will only be focusing on the case with defects in this paper and have adopted a simpler notation.}   

$\MM(n_{\rm m}, m_X; \{p_J,\vec{\nu}_J\})$ is a $4n_{\rm m}$-dimensional space that carries a natural Riemannian metric.  The geometry of $\MM$ will be discussed in the next subsection.  Here we note that the dimension can interpreted as follows.  A point in $\MM$ represents a nonlinear superposition of $n_{\rm m}$ ordinary monopoles in the presence of the defects.  Each monopole has four moduli associated to it: three for its position in physical three-space and a fourth whose conjugate momentum corresponds to an electric charge that each monopole can carry.  We let $R^n$, $n = 1,\ldots, 4n_{\rm m}$ denote local coordinates on $\MM$, and we write the family of solutions to the Bogomolny equation as
\begin{equation}\label{gensolution}
A_i = A_i(\vec{r}; R^n) ~, \qquad X = X(\vec{r}; R^n)~.
\end{equation}

Given a solution $(A_i,X)$ to the Bogomolny equation and a boundary value $Y_\infty$, there will be a unique solution to the secondary BPS equation, $D^i D_i Y + [X,[X,Y]] = 0$, for the $Y$ Higgs field \cite{Moore:2015szp}.  Hence, by \eqref{BPSeqns}, the electric field and therefore the electric charge, $\gamma_{\rm e}^{\rm phys}$, will be determined.  As we move in $\MM$, the electric charge will change.  In other words, the electric charge is a function on monopole moduli space, determined by $Y_\infty$.  Thus, if we fix an electric charge, solutions to \eqref{BPSeqns} carrying that charge lie in a subspace of the moduli space defined by a level set $\gamma_{\rm e}^{\rm phys} = \gamma_{\rm e}^{\rm phys}(R^n)$.

Next we turn to a discussion of moduli space geometry and monopole dynamics via the collective coordinate ansatz.  As we will explain, in this context it is more natural to proceed from solutions to the Bogomolny equation only, without imposing the secondary BPS equation for $Y$.  The effects of $Y$ will instead be felt through a potential energy on the moduli space, and possible values of the electric charge will be realized as constants of motion for the moduli space dynamcs.

\subsection{Moduli Space Geometry}

The kinetic terms in the action \eqref{Svan} for $(A_i, X)$ specify a metric on the infinite-dimensional space of field configurations through the identification of $\delta_t \equiv (\delta_t A_i, \delta_t X) = (\dot{A}_i, \dot{X})$ with a tangent vector at the point $(A_i,X)$.  This metric is the standard flat one:
\begin{equation}\label{bigmetric}
g(\delta_1, \delta_2) := \frac{1}{2\pi} \int_{\mathcal{U}} d^3 x \Tr \left\{ \delta_1 A_i \delta_2 A^i + \delta_1 X \delta_2 X \right\}~,
\end{equation}
and it induces a metric on the moduli space of singular monopoles.  (The factor of $1/(2\pi)$ turns out to be a convenient normalization; see \cite{Moore:2015szp} for details.) 

To determine the moduli space metric we need a set of tangent vectors to solutions $(A_i,X)$ of the Bogomolny equation that will generate motion along the moduli space.  These tangent vectors should therefore correspond to $(\delta A_i, \delta X)$, where $(A_i + \delta A_i, X + \delta X)$ is a solution to the Bogomolny equation linearized in the deformations, $\delta A_i, \delta X$.

By differentiating the Bogomolny equation with respect to the moduli $R^n$ appearing in \eqref{gensolution}, one finds that $(\pd_n A_i, \pd_n X)$, where $\pd_n \equiv \frac{\pd}{\pd R^n}$, solves the linearized equations.  However we must additionally require that the tangent vector $(\delta A_i, \delta X)$ be orthogonal to local gauge transformations, since the moduli space is the space of gauge-inequivalent solutions.  This is achieved by requiring $g(\delta, \delta_{\epsilon}) = 0$ for all $\epsilon$, where $\delta_{\epsilon} = (-D_\mu \epsilon, [\epsilon,X])$ is an infinitesimal gauge transformation.  The configuration $(\pd_n A_i, \pd_n X)$ can be adjusted to solve this constraint by shifting it by a local gauge transformation:
\begin{equation}\label{tangents}
\delta_n \equiv (\delta_n A_i, \delta_n X) := (\pd_n A_i - D_i \epsilon_n, \pd_n X + [\epsilon_n, X])~,
\end{equation}
with $\epsilon_n$ solving
\begin{equation}
D^i D_i \epsilon_n + [X, [X,\epsilon_n]] = D^i (\pd_n A_i)~.
\end{equation}
This condition ensures that $g(\delta_n, \delta_{\epsilon}) = 0$.  Then the components of the moduli space metric with respect to the local coordinates $R^n$ are
\begin{equation}\label{monopolemetric}
g_{mn} = \frac{1}{2\pi} \int_{\mathcal{U}} d^3 x \Tr \left\{ \delta_m A_i \delta_n A^i + \delta_m X \delta_n X \right\} ~.
\end{equation}

If one has an explicit family of solutions to the Bogomolny equation, \eqref{gensolution}, one can in principle compute the tangent vectors \eqref{tangents} and determine the metric directly from this definition.  This was carried out in \cite{Shah} for the family of solutions describing a single $\mathrm{SO}(3)$ monopole in the presence of defects studied in this paper.  Typically, however, in cases where the metric is known, it is obtained from other mathematical representations of the moduli space, and there is a great literature on the subject.  

Away from singular points the metric is hyperk\"ahler.  Co-dimension four singularities can exist in the moduli space of singular monopoles and are related to the phenomenon of monopole bubbling \cite{Kapustin:2006pk}, in which an 't Hooft defect emits or absorbs a smooth monopole, changing the asymptotic magnetic charge of the system.  In all known examples the singularities are of a fairly benign orbifold type.  Furthermore, if all 't Hooft defects are taken to be minimally charged, $|p_J| = 1$, then monopole bubbling does not occur, and the moduli space is smooth.

Another geometric construction on moduli space we will require is the Killing vector fields induced by global gauge transformations that preserve the Bogomolny equation and asymptotic data.  Such Killing fields are in one-to-one correspondence with the Cartan subalgebra, $\mathfrak{t}$, of the gauge group.  These vector fields generate isometries---in fact they are tri-holomorphic, generating isometries that preserve the hyperk\"ahler structure as well the metric.  The map
\begin{align}\label{Gmap}
\mathrm{G} :  \mathfrak{t} & \to \mathfrak{isom}(\MM) \cr
 H & \mapsto \mathrm{G}(H)~,
 \end{align}
 is constructed as follows.  Let $\epsilon_H$ be the unique solution to $D^i D_i \epsilon_H + [X,[X,\epsilon_H]] = 0$ with $\lim_{\vec{x} \to \infty} \epsilon_H = H$.  Then the Killing field $\mathrm{G}(H)$ is given at the point $(A_i,X)$ by the tangent vector $\delta_{\epsilon_H} = (-D_i \epsilon_H, [\epsilon_H X])$.  Since we are restricting to gauge group $G = \mathrm{SO}(3)$ in this paper, the Cartan subalgebra is one-dimensional and there will be a single linearly independent Killing field generated corresponding to the action of global gauge transformations.  Specifically, since $\exp(2\pi T_3) = \mathbbm{1} \in \mathrm{SO}(3)$, $\mathrm{G}(T^3)$ will be a Killing field that generates a $2\pi$-periodic isometry.
 
 \subsection{Collective Coordinate Dynamics on Monopole Moduli Space}\label{ssec:ccapprox}

The metric \eqref{monopolemetric} and Killing fields \eqref{Gmap} play a central role in the collective coordinate description of monopole dynamics.  The basic idea of the collective coordinate ansatz \cite{Manton:1981mp} is that time-dependent solutions of the full field theory describing monopole dynamics should be well-approximated by motion on moduli space---that is, by allowing the moduli to become functions of time (collective coordinates).  Intuitively, this should be true provided the collective coordinate velocities are small, since the moduli space is a minimum-energy surface in the space of field configurations.  Quantifying this condition requires some care.  

The reason the question is subtle is that field fluctuations around the monopole include modes of arbitrarily long wavelength for components of the fields along the $\mathfrak{u}(1)$ preserved by the vev.  In other words, there is no mass gap in the spectrum of fluctuations, and so energy can freely leak into radiation.  Nevertheless, radiation is sourced by accelerating charges and one might expect energy loss to be small if the time variation of the collective coordinates is small.  

In the context of classical time-dependent solutions in ordinary Yang--Mills--Higgs theory, the following mathematical result has been obtained by Stuart \cite{Stuart1994}.  Suppose the collective coordinates are slowly varying, such that time derivatives behave as $\pd_{t}^n R = O(\epsilon^n)$ for $n=1,2,3$, with $\epsilon$ a small parameter.  Given an initial field configuration that is close to the moduli space, such that the distance from the moduli space with respect to the metric \eqref{bigmetric} is $O(\epsilon^2)$, then the exact time-dependent solution to the Yang--Mills--Higgs equations will stay $O(\epsilon)$ close to a model trajectory, $R^{(0)}(t)$, for times $t \in [0,T]$ with $T=O(1/\epsilon)$ (in terms of the natural units of the problem set by the Higgs vev $m_X$).  The model trajectory is the geodesic on moduli space obtained by ignoring the coupling to radiation.

This result is consistent with a physical estimate of the energy lost to radiation over the same time scale.  Following \cite{Manton1988}, we evaluate the fields on the collective coordinate ansatz \eqref{gensolution} with the model trajectory $R^{(0)}(t)$ and consider the time-dependence of the asymptotic multipole expansion for the massless $\mathfrak{u}(1)$ components.  In general, since the monopole terms -- both magnetic and electric -- are time-independent, the leading contribution comes from dipole radiation.\footnote{In \cite{Manton1988}, which considered specific trajectories in the two-monopole system, there was additional suppression due to the fact that the magnetic dipole vanished, and the induced electric dipole moment was itself $O(\epsilon)$.}  The electric and magnetic dipole moments of the asymptotic fields can depend on the collective coordinates and are thus time-dependent.  Since dipole radiation has a total radiated power of order $P_{\rm rad} \sim (\pd_{t}^2 \vec{d} )^2$, where $\vec{d}$ is either the magnetic or electric dipole moment, (see \eg \cite{Landau:1982dva}), the rate of energy loss is $O(\epsilon^4)$.  One can view the effects of this energy loss as a radiation reaction force of $O(\epsilon^3)$ acting on the system.  Over time scales $T = O(1/\epsilon)$, one expects such a force to cause a deviation in the trajectory of $O(\epsilon)$, and this is consistent with the theorem in \cite{Stuart1994}.  

A third point of view on the limits of the collective coordinate approximation arises in the context of quantum Yang--Mills--Higgs.  In the semiclassical approximation to soliton states in quantum field theory, it is natural to take the collective coordinate velocities to be the same order as the Yang--Mills coupling, $g_{\rm ym}$, which is assumed to be small.  Hence $g_{\rm ym}$ plays the role of $\epsilon$ in the above discussion.  With this identification one ensures that quantum corrections from field fluctuations around the monopole are suppressed relative to the leading collective coordinate dynamics and can be treated perturbatively.  This perspective goes back to the original work on soliton quantization (see \eg \cite{Callan:1975yy,Gervais:1975pa} for the diagrammatic approach).  One can then define an effective Hamiltonian for the collective coordinates in the $n$-monopole sector by path-integrating out the field theoretic fluctuation fields around the background configuration \eqref{gensolution}.  

In the leading saddle-point approximation to this path integral, one is solving the classical equation of motion for the fluctuation field and inserting this solution back into the field theory action to arrive at an effective action for the collective coordinates.  This process can be carried out order by order in the small velocity expansion.  At zeroth order in time derivatives of the collective coordinates one finds the (classical) mass of the soliton, which is $O(g_{\rm ym}^{-2})$.  There are no terms at first order in time derivatives because the static soliton is an exact solution.  At second order in time derivatives, corresponding to $O(g_{\rm ym}^0)$ terms in the Hamiltonian, one recovers the standard two-derivative collective coordinate Hamiltonian, whose equations of motion reproduce the model trajectory $R^{(0)}(t)$.  One also recovers the first quantum correction to the soliton mass, which is independent of the collective coordinates.  The first effects of the coupling between collective coordinates and radiation modes enter the effective Hamiltonian for the collective coordinates at third order in time derivatives, corresponding to $O(g_{\rm ym})$.  It is these terms that provide the explicit radiation reaction force discussed above,\footnote{Explicit computations of these higher-derivative corrections have not been carried out, but an extension of the framework recently developed in \cite{Melnikov:2020ret,Melnikov:2020iol} to monopoles in Yang--Mills--Higgs theory would make it possible to do so.} and the same conclusion applies:  over time scales $T = O(1/\epsilon)$ these terms will lead to a deviation from the model trajectory of $O(\epsilon)$.    

All of these approaches consistently show that, in the slowly-varying regime, the trajectories resulting from Manton's collective coordinate approximation for Yang--Mills--Higgs theory remain $O(\epsilon)$ close to the true trajectories through times of $O(1/\epsilon)$.  The advantage of the third approach, based on the collective coordinate effective Hamiltonian, is that it has been extended to supersymmetric Yang--Mills--Higgs with the secondary Higgs field $Y$ and its relation to electric charge \cite{Gauntlett:1993sh,Sethi:1995zm,Gauntlett:1995fu,Lee:1996kz,Tong:1999mg,Bak:1999da,Bak:1999sv,Gauntlett:1999vc,Gauntlett:2000ks}, and with the inclusion of 't Hooft defects \cite{Moore:2015szp}.  We recall the key insights and results of this extension now.

As noted previously, the secondary BPS equation in \eqref{BPSeqns2}, and the equation $E_i = D_i Y$, imply that the electric charge depends on the vev $m_Y$ and the point in moduli space.  A analysis of this constraint shows that having configurations with electric charges $q \lesssim O(g_{\rm ym}^{-1})$ requires a hierarchy of scales $m_Y/m_X \sim O(g_{\rm ym})$.  Hence, one should treat $Y, A_0$ on the same footing as the collective coordinate velocities.  Specifically, one makes the ansatz
\begin{equation}\label{ccansatz}
A_i = A_i(\vec{r}; R^n(t))~, \qquad X = X(\vec{r}; R^n(t))~,
\end{equation}
and solves the remaining equations of motion for $A_0$ and $Y$ in this background, working perturbatively in $g_{\rm ym}$, under the assumption that
\begin{equation}\label{gscaling}
\dot{R}^n  \sim m_Y/m_X \sim O(g_{\rm ym})~.
\end{equation}
Upon inserting these expressions back into the action, one can integrate over space and, using the definition of the metric \eqref{monopolemetric}, one finds that the field theory action reduces to a particle mechanics action for the collective coordinates, $R^n(t)$.

This calculation was carried out in detail in \cite{Moore:2015szp} allowing for the presence of 't Hooft defects, and here we simply quote the results.  In fact, we will only give part of the results since we are not considering the dynamics of the fermionic degrees of freedom in this paper.  Index theory can be used to show that the fermions also carry $4 n_{\rm m}$ massless real degrees of freedom in the monopole background; these are the superpartners of the bosonic collective coordinates.  While this structure is essential for understanding the correct quantum mechanical model for the collective coordinates, it plays no role in the classical dynamics.  Hence, setting the fermionic degrees of freedom to zero, one finds the following expansion for the field theory Lagrangian \eqref{Stot} around the monopole background:
\begin{align}
L =&~ - \frac{4\pi}{\gymsq} \Tr(\gamma_{\rm m} X_\infty) + L_{\rm c.c.}^{(0)} + L_{\rm c.c.}^{(\thetaym)} + O(\gym)~, \qquad \textrm{where} \label{ccexpansion} \\
L_{\rm c.c.}^{(0)} =&~ \frac{4\pi}{\gymsq} \left[ \frac{1}{2} g_{mn} \left( \dot{R}^m \dot{R}^n - \mathrm{G}(Y_\infty)^m \mathrm{G}(Y_\infty)^n \right) \right] ~, \label{Lcc} \\
L_{\rm c. c.}^{(\thetaym)} =&~ \frac{\thetaym}{2\pi} \left[ \Tr(\gamma_{\rm m} Y_\infty) + g_{mn} (\dot{R}^m - \mathrm{G}(Y_\infty)^m) \mathrm{G}(X_\infty)^n \right] ~. \label{Lthetacc}
\end{align}

The terms in the first line, \eqref{ccexpansion}, are organized by scaling in $\gym$.  In units of the vev, $m_X$, the first term is $O(g_{\textrm{ym}}^{-2})$, $L_{\rm c.c.}^{(0)}$ is $O(1)$, and $L_{\rm c.c.}^{(\thetaym)}$ is $O(\gym)$.  Here we are using \eqref{gscaling}.  $L_{\rm c.c.}^{(0)}$ and $L_{\rm c.c.}^{(\thetaym)}$ are the bosonic pieces of two separate supersymmetry invariants identified in \cite{Moore:2015szp}.  The terms comprising $L_{\rm c.c.}^{(0)}$ form the bosonic part of a collective coordinate Lagrangian that was first obtained for monopoles without defects in supersymmetric Yang--Mills--Higgs theory in \cite{Gauntlett:1999vc,Gauntlett:2000ks}.  In particular, they feature a potential energy term given by the norm-squared of the Killing field $\mathrm{G}(Y_\infty)$.  Thus we see how the $Y$ Higgs field gives rise to a potential energy on the moduli space.  

The final term, $L_{\rm c.c.}^{(\thetaym)}$, and its fermionic completion, were first obtained in \cite{Moore:2015szp}.  This term is only dynamical in the presence of 't Hooft defects.  When defects are absent, the Killing field $\mathrm{G}(X_\infty)$ can be shown to be covariantly constant and $L_{\rm c.c.}^{(\thetaym)}$ becomes a total time derivative.  Since this term is $O(g_{\rm ym})$, we should either drop it or write all contributions to the effective Hamiltonian at this order, which include the first higher-derivative corrections to $L_{\rm c.c.}^{(0)}$.  In \cite{Moore:2015szp}, the focus was on certain BPS trajectories and their quantum analogs, where supersymmetry can be used to argue that the higher-derivative corrections are inessential.  

In this paper our interest is in generic collective coordinate motion, so we cannot make the same argument.  We will nevertheless keep the terms in $L_{\rm c.c.}^{(\thetaym)}$.  The reason is that, on the one hand, their effects are innocuous -- modifying the definition of the canonical momenta below and adding a correction to the parameter that controls the strength of the moduli space potential energy.  On the other hand, keeping these terms makes it easier to compare with \cite{Moore:2015szp}, where they are important for matching onto predictions from the Seiberg--Witten description of BPS states \cite{Seiberg:1994rs}.\footnote{One can always choose to set $\theta_{\rm ym} = 0$ above and in the following.  Then \eqref{ccexpansion} is consistent with a strict $g_{\rm ym}$ expansion.}

In particular, we recognize the constant term in $L_{\rm c.c.}^{(\thetaym)}$ as the remaining magnetic-charge contribution to the BPS mass, \eqref{BPSmass}, once $\gamma_{\rm e}^{\rm phys}$ is expressed in terms of $\gamma_{\rm m}$ and $\gamma_{\rm e}$ using \eqref{echargeshift}.  The $\gamma_{\rm e}$ contribution will instead be obtained from conserved momenta in the collective coordinate dynamics.  We set
\begin{equation}
M_{\gamma_{\rm m}} = \frac{4\pi}{\gymsq} \Tr(\gamma_{\rm m} X_\infty) - \frac{\thetaym}{2\pi} \Tr (\gamma_{\rm m} Y_\infty)~,
\end{equation}
and we find that the Lagrangian \eqref{ccexpansion} leads to the conjugate momenta
\begin{equation}\label{momenta}
\pi_m = \frac{4\pi}{\gymsq} g_{mn} \dot{R}^n + \frac{\thetaym}{2\pi} g_{mn} \mathrm{G}(X_\infty)^n~,
\end{equation}
and collective coordinate Hamiltonian
\begin{align}
H_{\rm c.c.} =&~ M_{\gamma_{\rm m}} + \frac{\gymsq}{4\pi} \cdot \half \left( \pi_m - \tfrac{\thetaym}{2\pi} \mathrm{G}(X_\infty)_m \right) g^{mn} \left( \pi_n - \tfrac{\thetaym}{2\pi} \mathrm{G}(X_\infty)_n \right) + \cr
&~ +  \frac{4\pi}{\gymsq} \cdot \half g_{mn} \mathrm{G}(Y_\infty)^m \mathrm{G}(Y_\infty)^n + \frac{\thetaym}{2\pi} g_{mn} \mathrm{G}(X_\infty)^m \mathrm{G}(Y_\infty)^n ~.
\end{align}
Note that the momenta are $O(g_{\rm ym}^{-1})$ since the velocities are $O(g_{\rm ym})$ while the mass is $O(g_{\rm ym}^{-2})$.  We note that the Hamiltonian can also be written in the form
\begin{align}\label{ccHam}
H_{\rm c.c.} =&~ M_{\gamma_{\rm m}} + \frac{g_{\rm ym}^2}{4\pi} \cdot \half \left[ \pi_m g^{mn} \pi_n + \mathrm{G}(\mathcal{Y}_{\infty})^m g_{mn} \mathrm{G}(\mathcal{Y}_{\infty})^n  \right] - \tilde{\theta}_{\rm ym} \pi_n \mathrm{G}(X_\infty)^n ~,
\end{align}
where we have introduced the combination
\begin{equation}
\mathcal{Y}_{\infty} = \frac{4\pi}{g_{\rm ym}^2} Y_\infty + \frac{\theta_{\rm ym}}{2\pi} X_\infty ~,
\end{equation}
and used the linearity of the $\mathrm{G}$-map.  The momenta and Hamiltonian are subject to $O(1)$ and $O(\gym)$ corrections, respectively, coming from higher-derivative terms.  As discussed above, these higher derivative terms originate from the coupling of the collective coordinates to radiation modes in the full field theory.

Thus, under the scaling assumptions \eqref{gscaling}, the coupling to radiation continues to be suppressed as it is in ordinary Yang--Mills--Higgs theory.  This strongly suggests there should exist a direct analog of Stuart's theorem \cite{Stuart1994} in the supersymmetric context, with or without 't Hooft defects, for the moduli space with potential approximation.  For such an extension of the theorem, the collective coordinate ansatz for all of the fields in the presence of 't Hooft defects would be the one given in subsection 4.3.1 of \cite{Moore:2015szp}.   One does not expect the presence of defects to cause additional difficulties in the analysis since the linearized fluctuation operator controlling the radiation spectrum is sufficiently regular at the defect points: no special boundary conditions are required, and the modes are locally $L^2$ in a neighborhood of the defect points.  Indeed, this was a key point in the analysis of \cite{Moore:2014jfa} determining the dimension of the moduli space from a Callias index theorem.\footnote{No choice of self-adjoint extension is needed as it was in \cite{Kazama:1976fm,Goldhaber:1977xw,Callias:1977cc}.  The difference between those references and the situation considered in \cite{Moore:2014jfa} is that the background Higgs field on which the linearized fluctuation operator depends also has a singularity in the presence of an 't Hooft defect. This leads to a cancelation in the leading singularity of the operator analyzed in the earlier references.}

The same procedure of solving the equations of motion perturbatively, as described under \eqref{ccansatz}, leads to an expression for the electric charge, $\gamma_{\rm e}$, as function on moduli space \cite{Tong:1999mg,Moore:2015szp}:
\begin{equation}\label{electriccharge}
\gamma_{\rm e} =  q T^3 ~, \qquad q =  \half \mathrm{G}(T^3)^m \pi_m ~.
\end{equation}
In the semiclassical quantization of the collective coordinate dynamics, $q$ is constrained to take integer values, since $\mathrm{G}(T^3)^m$ generates a $2\pi$-periodic isometry and the corresponding momentum eigenvalues are quantized.\footnote{For the supersymmetric Yang--Mills--Higgs theory discussed here, $q$ would take on only even integer values due to the fact that all fields transform in the adjoint representation of the gauge group.  In the notation of \cite{Moore:2015szp}, $q = -2n_{\rm e}$.}  In the classical theory, however, $q$ can be any real number.  Since the momenta are $O(g_{\rm ym}^{-1})$ in the scaling regime we work in, it is natural to consider charges $q \lesssim O(g_{\rm ym}^{-1})$.  Note such charges still lead to an electric field that is $O(g_{\rm ym})$ according to \eqref{asymptotics} and hence suppressed compared to the magnetic field.

In the remainder of this paper we will analyze a class of solutions to the Bogomolny equation, describing one smooth monopole in the presence of any number of 't Hooft defects.  We will construct simulations of monopole-defect interactions based on the corresponding moduli space geometry and collective coordinate Hamiltonian.

\section{Simulating One Monopole and $k$ Defects}\label{sec:BCD}

\subsection{The BCD solutions}

The solutions presented here were first obtained in \cite{Cherkis:2007jm,Cherkis:2007qa} using a form of the Nahm transform \cite{Nahm:1979yw} for singular monopoles developed in \cite{Cherkis:1997aa}.  Later, the solutions were recovered in \cite{Blair:2010kz} from a modified Nahm transform referred to as the bow construction and developed in \cite{Cherkis:2010bn,Blair:2010vh}.

As above, $\vec{r}$ denotes the general position vector in $\mathbbm{R}^3$, $\vec{\nu}_J$ are the positions of the singularities indexed by $J = 1,2,\ldots,k_{\rm t}$, and $\vec{r}_J \equiv \vec{r} - \vec{\nu}_J$.  We let $k_{\rm t}$ denote the total number of 't Hooft defects.  Each defect is taken to be minimally charged, $|p_J| = 1$, since a non-minimally charged singularity can be obtained by letting some of the $\vec{\nu}_J$ coincide.  We denote the moduli corresponding to the smooth monopole's position by $\vec{R}$.  It will also be convenient to define $\vec{R}_J \equiv \vec{R} - \vec{\nu}_J$ as the smooth monopole position relative to the $J^{\rm th}$ defect and $\vec{z} \equiv \vec{r} - \vec{R}$ as the observation point relative to the smooth monopole.\footnote{To compare with \cite{Cherkis:2007qa}, let $\vec{r} \to \vec{t}$, $\vec{r}_J \to \vec{t}_J$, $\vec{R} \to -\vec{T}$, $\vec{R}_J \to - \vec{T}_J$.  Additionally, due to a different normalization convention for the generators $T^a$, $(A_{ai},X_i)^{\textrm{here}} = 2 (A_{ai},X_a)^{\textrm{there}}$.  For the same reason, $m_{X}^{\textrm{here}} = 2\lambda^{\textrm{there}}$, and we set $v^{\textrm{here}} = 2z \alpha^{\textrm{there}}.$}  These vectors are not all independent; in particular, $\vec{r}_J - \vec{R}_J = \vec{z}$, for each $J$.  The same letter without the arrow notation always denotes the magnitude of the vector: $R_J = |\vec{R}_J|$, \etc  The following quantities appear regularly in the following and are given special names:
\begin{align}
& \mathcal{P}_J := \sqrt{ (r_J + R_J)^2 - z^2} = \sqrt{ 2 (r_J R_J + \vec{r}_J \cdot \vec{R}_J)}~, \qquad v := \half \sum_{J=1}^{k_{\rm t}} \ln\left( \frac{r_J + R_J + z}{r_J + R_J - z}\right) ~.
\end{align}

In terms of these quantities, the Blair--Cherkis--Durcan solutions are
\begin{align}\label{BCDsolution}
A_{ia} =&~ \epsilon_{iaj} z^j f_0 + \sum_{J=1}^{k_{\rm t}} \epsilon_{ijk} r_{J}^k \left \{ z_a R_{J}^j f_J - (z^2 \delta_{a}^{~j} - z_a z^j)  g_J \right\} ~, \cr
X_a =&~ z_a g_0 + \sum_{J=1}^{k_{\rm t}} (z^2 \delta_{aj} - z_a z_j) R_{J}^j g_J ~,
\end{align}
with the functions $f,g$ given by
\begin{align}\label{fandgs}
f_0 =&~ \frac{1}{z} \left\{ \left( m_X + \sum_{J=1}^{k_{\rm t}} \frac{(r_J + R_J)}{\PP_{J}^2} \right)\csch{(m_X z + v)} - \frac{1}{z} \right\}~, \cr
f_J =&~ \frac{1}{z r_J \PP_{J}^2} \coth{(m_X z + v)}~, \cr
g_0 =&~ \frac{1}{z} \left\{ \frac{1}{z} - \left( m_X + \sum_{J=1}^{k_{\rm t}} \frac{1}{2 r_J} \right) \coth{(m_X z + v)} \right\}~, \cr
g_J =&~ \frac{1}{z r_J \PP_{J}^2} \csch{(m_X z + v)} ~.
\end{align}
The solutions are written in a hedgehog-type gauge, where spatial directions indexed by $i,j = 1,2,3$ are correlated with directions in the Lie algebra indexed by $a,b = 1,2,3$.  We use Einstein summation conventions for repeated indices of type $i,j$ and type $a,b$, but we always write the sum over defects explicitly.  $\epsilon_{ijk}$ is the totally antisymmetric symbol with $\epsilon_{123} = 1$.  In the limit where all defects are sent to infinity, $\nu_J \to \infty$, one sees that $v \to 0$ and the terms involving $f_J, g_J$ vanish in \eqref{BCDsolution}.  Furthermore, the terms from the sums over $J$ in $f_0,g_0$ vanish, and $(A_{ai},X_a)$ reduces to the Prasad--Sommerfield solution for the smooth monopole.  As we approach the $J^{\rm th}$ defect, the leading singularity of the Higgs field is evident from the $1/(2r_J)$ term in $g_0$.  

Meanwhile, the asymptotic behavior of the Higgs field as $r \to \infty$ can be extracted from the $g_0$ term:
\begin{equation}
X = - \left( m_X - \frac{(2 - k_{\rm t})}{2r}  \right) \hat{z}_a T^a + o(1/r)~.
\end{equation}
By making patchwise gauge transformations on the two-sphere at infinity, $-\hat{z}_a T^a$ can be rotated to $T^3$.  Comparing with the asymptotic form, \eqref{asymptotics}, \eqref{magcharge}, we see that $n_{\rm m} =1$.  Hence this solution represents a single smooth monopole in the presence of the defects, as advertised.

The full family of solutions with $n_{\rm m} = 1$ depends on four moduli.  The solution \eqref{BCDsolution} exhibits dependence on three of these parameters, $\vec{R} \in \mathbbm{R}^3$, corresponding to the smooth monopole's position.  Dependence on the fourth modulus, $R^4$, can be implemented by acting on the configuration $(A_{ai},X_a)$ with an asymptotically nontrivial gauge transformation, $\cg = \exp(R^4 \epsilon_{X_\infty})$, where $\epsilon_{X_\infty}$ is the unique solution to $D^i D_i \epsilon_{X_\infty} + [X,[X,\epsilon_{X_\infty}]] = 0$ satisfying $\lim_{r \to \infty} \epsilon_{X_\infty} = X_\infty$ and regular in the interior.  $R^4$ is a circle coordinate since the gauge group is compact.    We will not need to carry this out explicitly, however.  The reason is that the local energy density in the fields is a gauge-invariant quantity and therefore will be independent of $R^4$.  We turn to the computation of the energy density next.

\subsection{Magnetic Field and Energy Density}

Since we are treating the effects of the second Higgs field and electric charge as a perturbation, we only consider the leading order contribution to the energy density, \eqref{endenfull}, due to the magnetic and primary Higgs field:
\begin{equation}
\mathcal{E}_{\rm m} := \Tr \left\{ B_i B^i + D_i X D^i X \right\} = (D_i X)_a (D^i X)^a = B_{ia} B^{ia}~,
\end{equation}
where in the second and third steps we used the Bogomolny equation, $B_i = D_i X$.  In this subssection we outline the computation of the components of the magnetic field, $B_{ai} = (D_i X)_a = \pd_i X_a + \epsilon_{abc} A_{i}^b X^c$, and the magnetic energy density, $\mathcal{E}_{\rm m}$, for the BCD solutions.

The derivative $\pd_i \equiv \frac{\pd}{\pd r^i}$ acts on $\vec{z}$ and the $\vec{r}_J$, whereas $\vec{R}_J$ is a constant.  Thus, for example,
\begin{equation}
\pd_i g_0 = \frac{\pd g_0}{\pd z} \frac{z_i}{z} + \sum_{K=1}^{k_{\rm t}} \frac{\pd g_0}{\pd r_K} \frac{r_{K i}}{r_K} ~.
\end{equation}
The partial derivatives of the $f$'s and $g$'s with respect to $z$ and $r_K$ can be straightforwardly evaluated, but we suppress it here.  

Our main focus in the computation is to express $\pd_i X_a + \epsilon_{abc} A_{i}^b X^c$ in a minimal set of tensor structures for the free $a$ and $i$ indices, and determine the scalar functions multiplying each of those tensor structures.  We use the identities
\begin{align}
\epsilon_{ijk} \epsilon^{ibc} =&~ \delta_{j}^{~b} \delta_{k}^{~c} - \delta_{j}^{~c} \delta_{k}^{~b} ~, \cr
\epsilon_{ijk} \epsilon^{abc} =&~ \delta_{i}^{~a} \delta_{j}^{~b} \delta_{k}^{~c} + \delta_{i}^{~b} \delta_{j}^{~c} \delta_{k}^{~a} + \delta_{i}^{~c} \delta_{j}^{~a} \delta_{k}^{~b}  - \delta_{i}^{~b} \delta_{j}^{~a} \delta_{k}^{~c} - \delta_{i}^{~c} \delta_{j}^{~b} \delta_{k}^{~a} - \delta_{i}^{~a} \delta_{j}^{~c} \delta_{k}^{~b} ~,
\end{align}
to eliminate all $\epsilon$ symbols.  A minimal set of tensor structures can be taken as
\begin{equation}\label{tensorstructures}
\textrm{tensor structures:} \quad \delta_{ia}~, ~~z_i z_a~, ~~z_i R_{Ja}~, ~~ R_{Ji} z_a~, ~~R_{Ji} R_{Ka} ~,
\end{equation}
since we can use $\vec{r}_J = \vec{z} + \vec{R}_J$ to eliminate all appearances of $r_{Ji}$ and $r_{Ja}$.

The coefficient functions associated with the tensor structures \eqref{tensorstructures} are denoted $h$, $h_{00}$, $h_{0J}$, $h_{J0}$, and $h_{JK}$, respectively, so that the magnetic field is
\begin{equation}\label{Bviahs}
B_{ia} = h \delta_{ia} + h_{00} z_i z_a + \sum_{J=1}^{k_{\rm t}} (h_{0J} z_i R_{Ja} + h_{J0} R_{Ji} z_a) + \sum_{J,K=1}^{k_{\rm t}} h_{JK} R_{Ji} R_{Ka} ~.
\end{equation}
Tedious but straightforward computation yields
\begin{align}\label{hfunctions}
h =&~ g_0 (1 + z^2 f_0) - \sum_J ( \vec{z} \cdot \vec{R}_J + z^2 \vec{z} \cdot \vec{r}_J g_0 ) g_J + \cr
&~ + z^2 \sum_{J,K} (  \vec{z} \cdot \vec{R}_J \vec{z} \cdot \vec{R}_K - z^2 \vec{R}_J \cdot \vec{R}_K ) f_J g_K ~, \cr
h_{00} =&~ \frac{1}{z} \frac{\pd g_0}{\pd z} - f_0 g_0 + \sum_J \left( \frac{1}{r_J} \frac{\pd g_0}{\pd r_J} - \frac{\vec{z} \cdot \vec{R}_J}{z} \frac{\pd g_J}{\pd z} + (z^2 g_0 + \vec{z}\cdot \vec{R}_J f_0) g_J \right) + \cr
&~ - \sum_{J,K} \left( \frac{ \vec{z} \cdot \vec{R}_J}{r_K} \frac{\pd g_J}{\pd r_K} + z^2 (\vec{r}_J \cdot \vec{R}_K g_J g_K - \vec{R}_J \cdot \vec{R}_K f_J g_K) \right) ~, \cr
h_{0J} =&~ (2 + z^2 g_0) g_J + z \frac{\pd g_J}{\pd z} + z^2 \sum_K \left( \frac{1}{r_K} \frac{\pd g_J}{\pd r_K} - \vec{z} \cdot \vec{R}_K f_J g_K \right) ~, \cr
h_{J0} =&~ - (1 + z^2 f_0) g_J + \frac{1}{r_J} \frac{\pd g_0}{\pd r_J} - \sum_K \left( \frac{ \vec{z} \cdot \vec{R}_K}{r_J} \frac{\pd g_K}{\pd r_J} + z^2 g_J (\vec{z} \cdot \vec{R}_K f_K -  \vec{z} \cdot \vec{r}_K g_K ) \right)~, \cr
h_{JK} =&~ z^2 \left( \frac{1}{r_J} \frac{\pd g_K}{\pd r_J} + z^2 g_J f_K \right) ~.
\end{align}
The energy density of the BCD solution is obtained by squaring \eqref{Bviahs}:
\begin{align}\label{endenfinal}
\mathcal{E}_{\rm m} =&~ 3 h^2 + z^4 h_{00}^2 + 2 z^2 h h_{00} + 2 \sum_{I=1}^{k_{\rm t}} \vec{z} \cdot \vec{R}_I (h + z^2 h_{00}) (h_{0I} + h_{I0}) + \cr
&~ + \sum_{I,J=1}^{k_{\rm t}} \left\{ \vec{R}_I \cdot \vec{R}_J \left[ z^2 (h_{0I} h_{0J} + h_{I0} h_{J0}) + 2 h h_{IJ} \right] + \right. \cr
&~ \qquad \qquad \qquad \qquad \left. + 2 \vec{z} \cdot \vec{R}_I \vec{z} \cdot \vec{R}_J (h_{00} h_{IJ} + h_{0I} h_{J0}) \right\} + \cr
&~ + 2 \sum_{I,J,K=1}^{k_{\rm t}} \vec{z} \cdot \vec{R}_K \vec{R}_I \cdot \vec{R}_J (h_{I0} h_{JK} + h_{0I} h_{KJ})  + \cr
&~ +  \sum_{I,J,K,L=1}^{k_{\rm t}} \vec{R}_I \cdot \vec{R}_K \vec{R}_J \cdot \vec{R}_L h_{IJ} h_{KL} ~.
\end{align}
Through the formulae of this section one thus obtains $\mathcal{E}_{\rm m}$ as a function of $\vec{r} \in \mathbbm{R}^3$, the $3(1 + k_{\rm t})$ parameters $\{\vec{R}, \vec{\nu}_J \}$, and the mass scale $m_X$.  As is clear from \eqref{fandgs}, $m_{X}^{-1}$ sets the natural length scale for the field configuration.

\begin{figure}[th!]
\begin{center}
\includegraphics[width=8.0cm]{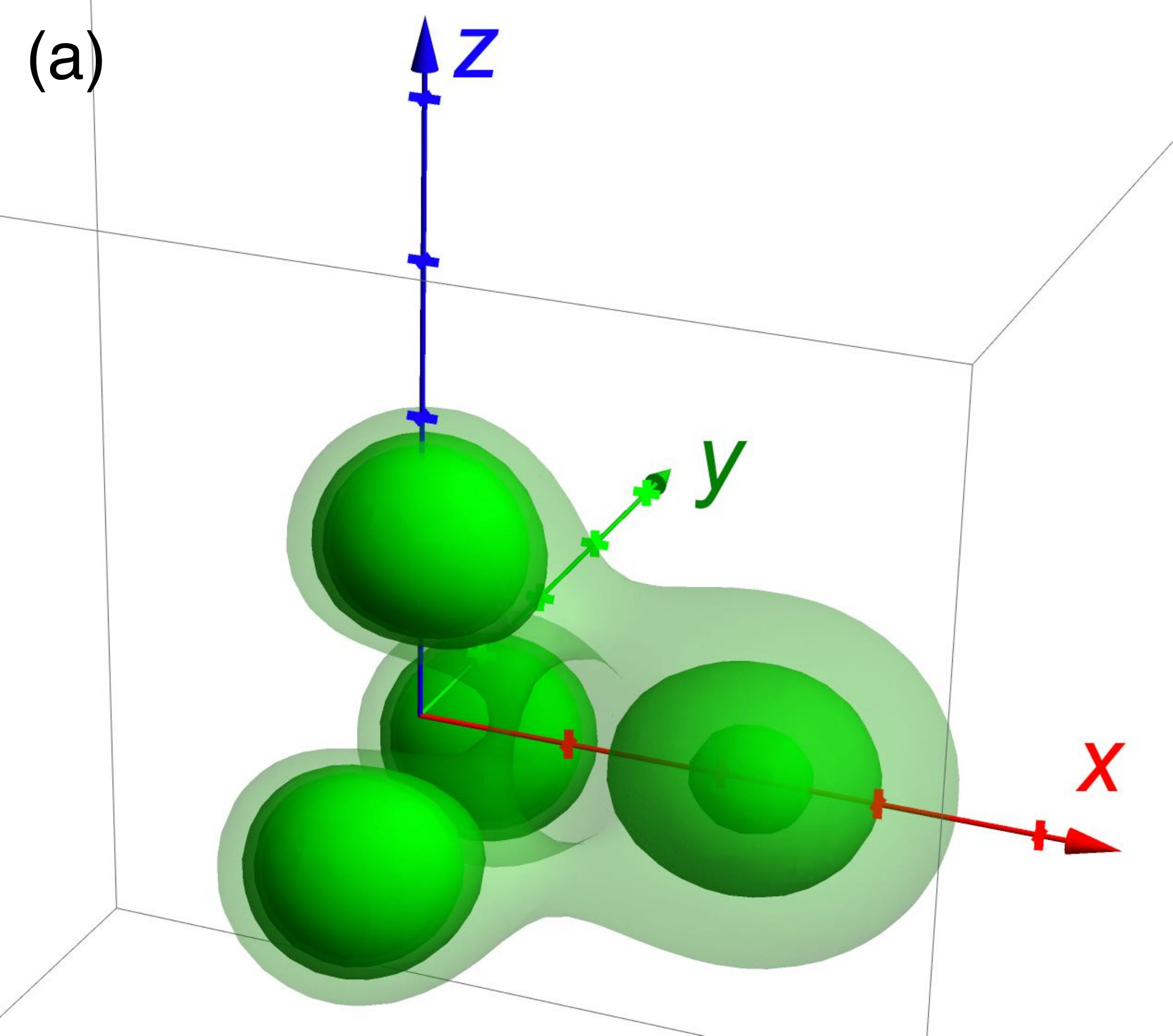} \\ \includegraphics[width=8.0cm]{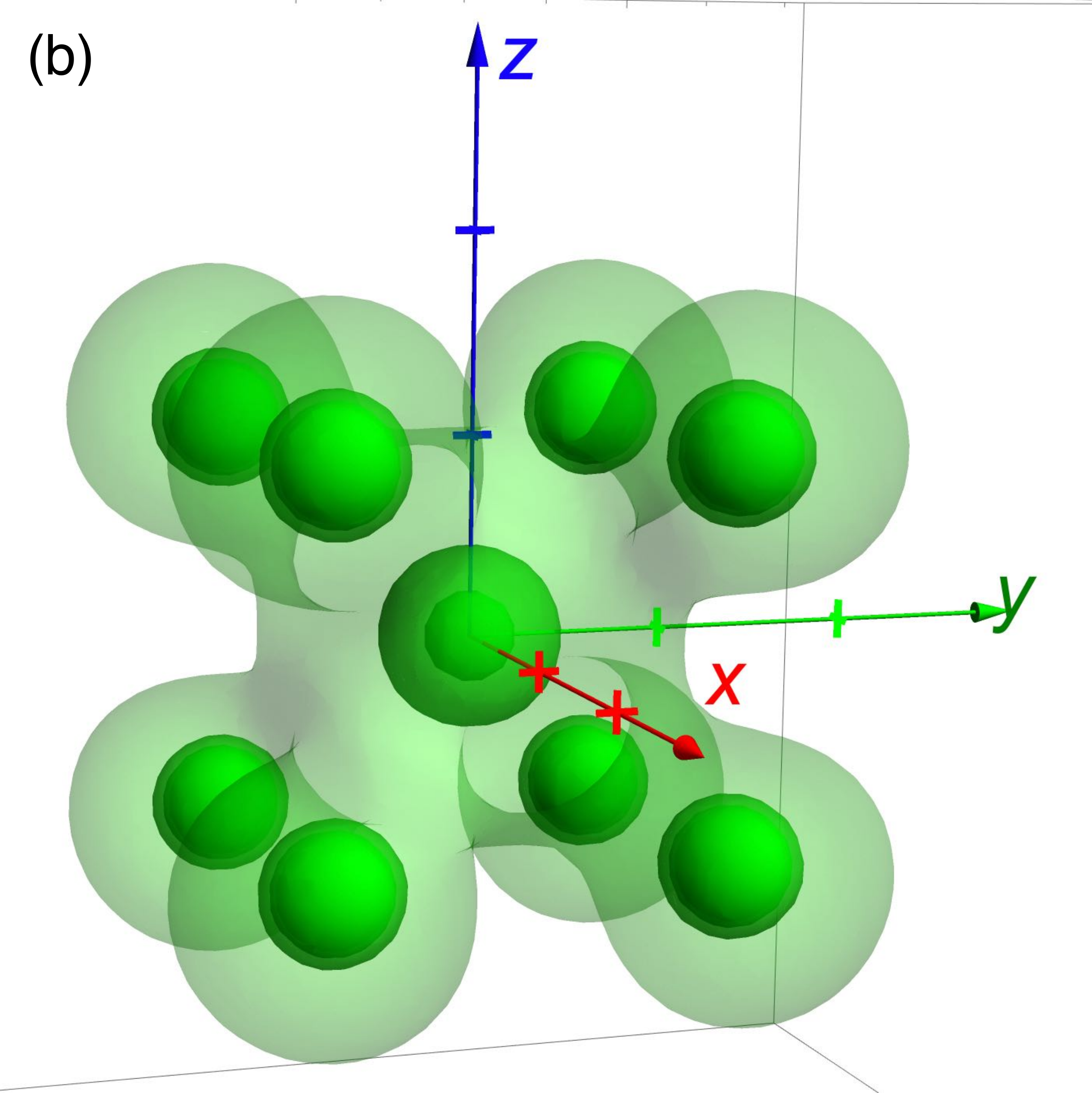}
\end{center}
\caption{Energy density plots.  The crosses on the axes are spaced one unit of $m_{X}^{-1}$ apart.  The top configuration, (a), has three defects at the vertices of an equilateral triangle in the $y$-$z$-plane, centered on the origin with sides of length 2, while the position of the smooth monopole is taken to be $(2.5,0,0)$.  The level sets of the energy density, \eqref{endenfinal}, for the top configuration are $0.55 + 0.63 n$, $n=0,1,2,3$, in units of $m_X$.  The bottom configuration, (b) has eight defects placed on the vertices of a cube with sides of length 2, centered at the origin, and the smooth monopole at the origin.  The values of the energy density are $0.88 +15 n$, $n=0,1,2,3$.}
\label{fig:EnergyPlots}
\end{figure}

We have constructed a module in Mathematica that takes as input $1 + k_{\rm t}$ vectors, $\{ \vec{R}, \vec{\nu}_1, \vec{\nu}_2, \ldots \}$, in units of $m_{X}^{-1}$, and produces a three-dimensional plot of several level sets of $\mathcal{E}_{\rm m}$, utilizing Mathematica's RegionPlot3D.  The value of the energy density for each surface increases in regular steps, and we have correlated these values with the opacity of the surface, with the lowest value corresponding to the most transparent surface and the highest value corresponding to a completely opaque surface.  This allows one to see ``inside'' the configuration.  See Figure \ref{fig:EnergyPlots} for two examples.

The code determines the values of the energy density to use based on a pre-sampling of values for the requested configuration.  It attempts to ensure that the local maximum at the core of the smooth monopole lies between the values for the third and fourth surface, so that the smooth monopole remains semi-transparent.  This however will not be possible if the smooth monopole is too close to a defect, such that there is not a well-isolated local maximum corresponding to its position.  The energy density has a $1/r_{J}^4$ singularity as one approaches the $J^{\rm th}$ defect, so the defects will always be accumulation points for the surfaces.

The code is denoted ``EnergyPlot'' in the Mathematica notebook included with the ancillary materials of this submission.  In addition to the position vector of the smooth monopole and a list (of arbitrary length) of position vectors for the defects, the code takes three further arguments:   the number of initial plot points to use in the argument of Mathematica's RegionPlot3D, the size of the final image in pixels, and the position from which to view the configuration.  (The output, however, can be rotated at will within Mathematica.)  The examples in Figure \ref{fig:EnergyPlots} used a relatively high value of 60 plot points and took 16 minutes and 5.5 hours respectively to render.  In general, we expect the computation time to scale like $k_{\rm t}^4$ due to the quartic term in the last line of \eqref{endenfinal}.

Somewhat faster computations might be possible, especially for high $k_{\rm t}$ values, by utilizing the identity $\Tr(D_i X D^i X) = \pd_i \pd^i \Tr(X^2)$.  The approach we have presented was motivated in part by the desire to have explicit expressions available that could be used to visualize the magnetic field itself.

\subsection{Motion on Multi-centered Taub-NUT}

The motion of the smooth monopole in the presence of the defects is determined by the equations of motion following from the collective coordinate Hamiltonian, \eqref{ccHam}.  This Hamiltonian requires the input of the metric on moduli space, $g_{mn}$, and the tri-holomorphic Killing vector fields $\mathrm{G}(X_\infty)^m,\mathrm{G}(\mathcal{Y}_\infty)^n$.  

The moduli space is the $k_{\rm t}$-centered Taub--NUT manifold \cite{Cherkis:1997aa}, and the metric is known explicitly.  Let $H(\vec{R})$ be the harmonic function on $\mathbbm{R}^3 \setminus \{ \vec{\nu}_J \}$ given by 
\begin{equation}
H(\vec{R}) = 1 + \frac{1}{2 m_X} \sum_{J=1}^{k_{\rm t}} \frac{1}{ |\vec{R} - \vec{\nu}_J|} ~,
\end{equation}
with $| \cdot |$ the standard Euclidean norm.\footnote{In this subsection we use the notation $\pd_i \equiv \frac{\pd}{\pd R^i}$.  There should be no confusion as the coordinates on the physical space, $\vec{r}$, do not appear in this subsection.}  Let $\Omega_i(\vec{R})$ be a $\mathrm{U}(1)$ gauge field on $\mathbbm{R}^3 \setminus \{ \vec{\nu}_J \}$ determined by
\begin{equation}\label{selfdual}
\pd_i \Omega_j - \pd_j \Omega_i = \epsilon_{ij}^{~~k} \pd_k H~.
\end{equation}
Here, as always, the flat Euclidean metric is used to raise/lower indices of type $i,j$.  Then the metric on the four-dimensional multi-centered Taub--NUT space takes the form
\begin{align}
g_{mn} dR^m dR^n =&~ m_X H \delta_{ij} dR^i dR^j + m_X H^{-1} \left( \frac{1}{2m_X} dR^4 + \Omega_i dR^i \right)^2~.
\end{align}
Here $R^4$ is a circle coordinate with periodicity $R^4 \sim R^4 + 4\pi$, and the manifold restricts to a circle bundle over $\mathbbm{R}^3 \setminus \{ \vec{\nu}_J \}$.  As $R \equiv |\vec{R}| \to \infty$, the size of the circle remains finite, and the overall normalization of the metric can be fixed by comparing to the definition \eqref{monopolemetric} in this limit \cite{Shah,Moore:2015szp}.  As one approaches a nut point, $\vec{R} \to \vec{\nu}_J$, the circle fiber shrinks to zero size such that the total space is smooth.

The vector field $\pd_4 \equiv \frac{\pd}{\pd R^4}$ that generates motion along the circle fiber is, up to rescaling, the only tri-holomorphic Killing field.  It follows from the periodicity of $R^4$ that $\mathrm{G}(T^3) = 2 \pd_4$.  Hence by linearity of the $\mathrm{G}$-map and \eqref{Higgsvevs},
\begin{equation}
\mathrm{G}(X_\infty) = 2 m_X \pd_4 ~, \qquad \mathrm{G}(\mathcal{Y}_\infty) = 2 m_{\YY} \pd_4 \equiv 2 \left(\frac{4\pi m_Y}{g_{\rm ym}^2} + \frac{\theta_{\rm ym} m_X}{2\pi} \right) \pd_4~.
\end{equation}
We then find the conjugate momenta and Hamiltonian from \eqref{momenta} and \eqref{ccHam} to be 
\begin{align}
\pi_4 =&~ \half H^{-1} \left[ \frac{1}{2m_X} \left( \frac{4\pi}{g_{\rm ym}^2} \dot{R}^4 + \frac{\theta_{\rm ym}}{2\pi} (2m_X) \right) + \frac{4\pi}{g_{\rm ym}^2} \dot{R}^i \Omega_i  \right]~, \cr
\pi_i =&~ \frac{4 \pi}{g_{\rm ym}^2} m_X H \dot{R}_i + 2 m_X \pi_4 \Omega_i ~,
\end{align}
and
\begin{align}
H_{\rm c.c.} =&~ \frac{g_{\rm ym}^2}{4\pi} m_X \bigg\{ \half m_{X}^{-2} H^{-1} \left( \pi_i - 2m_X \pi_4 \Omega_i \right) \delta^{ij}  \left( \pi_j - 2m_X \pi_4 \Omega_j \right) + \cr
&~ \qquad \qquad + 2 \pi_4 \left( H \pi_4 - \frac{\theta_{\rm ym}}{2\pi} \right) + \half \left( \frac{m_{\YY}}{m_X} \right)^2 H^{-1} \bigg\} ~.
\end{align}
The $m_{\YY}$ term provides a potential energy well in the vicinity of each defect, which can lead to bound motion.  Meanwhile $\pi_4$ is the electric charge $q$, \eqref{electriccharge}:
\begin{equation}
q =  \pi_4 ~.
\end{equation}

Since $m_{X}^{-1}$ sets the natural length scale in the physical $\mathbbm{R}^3$, we work with dimensionless position variables $\RR^i = m_{X} R^i$ and parameters $\tilde{\nu}_{Ji} = m_{X} \nu_{Ji}$ so that
\begin{equation}
H = 1 + \frac{1}{2} \sum_{J = 1}^{\rm k_{\rm t}} \frac{1}{ \left| \vec{\RR} - \vec{\tilde{\nu}}_J \right| }~.
\end{equation}
We note that $R^4$ is already dimensionless.  We also define a dimensionless time, dimensionless momenta, and a dimensionless parameter $C$ according to
\begin{equation}\label{dimless}
\tau = \frac{g_{\rm ym}}{2\sqrt{\pi}} m_X t~, \qquad \tilde{\pi}_i = \frac{g_{\rm ym}}{2\sqrt{\pi}} m_{X}^{-1} \pi_i ~, \qquad \qq = \frac{g_{\rm ym}}{2\sqrt{\pi}} q~,  \qquad C = \frac{g_{\rm ym} m_{\YY}}{4\sqrt{\pi} \, m_X} ~.
\end{equation}
The factors of $g_{\rm ym}$, together with \eqref{gscaling}, ensure that $\pd_\tau \tilde{R}^n$, $\tilde{\pi}_i, \qq$, and $C$ are all naturally $O(1)$ quantities.  The factors of $2$ and $\pi$ are for convenience.  

The expressions for the conjugate momenta now take the form
\begin{align}
\qq =&~ \half H^{-1} \left[ \half (\pd_\tau \tilde{R}^4) + \frac{g_{\rm ym} \theta_{\rm ym}}{4\pi^{3/2}} + (\pd_\tau \tilde{R}^i) \Omega_i \right] ~, \cr
\tilde{\pi}_i =&~ H \pd_\tau \tilde{R}_i + 2 \qq \Omega_i ~,
\end{align}
while the dynamical equations are
\begin{align}
\pd_\tau \qq = \frac{1}{m_X} \dot{p}_4 =&~ - \frac{1}{m_X}  \frac{\pd H_{\rm c.c.}}{\pd R^4} = 0~, \cr
\pd_\tau \tilde{\pi}_i  = \frac{1}{m_{X}^2} \dot{\pi}_i =&~ - \frac{1}{m_X} \frac{\pd H_{\rm c.c.}}{\pd \RR^i} \cr
=&~ \frac{2 \qq}{H} (\tilde{\pi}^j + 2 \qq \Omega^j) (\pd_{\RR^i} \Omega_j) + \left[ \frac{ (\tilde{\pi}_j + 2 \qq \Omega_j)^2}{2 H^2} + \frac{2 C^2}{H^2} - 2 \qq^2 \right] \pd_{\RR^i} H ~. \qquad
\end{align}
From the first equation we learn that the electric charge, $\qq$, is a constant of motion.  The remaining equations determine the motion of the smooth monopole on $\mathbbm{R}^3$.  These equations are more conveniently expressed in terms of the shifted momentum variables
\begin{equation}\label{newmomentum}
\pp_i := \tilde{\pi}_i  -2\qq \Omega_i = H \pd_{\tau} \RR_i ~,
\end{equation}
which leads to
\begin{equation}\label{Newton}
\pd_\tau \pp_i =  \frac{2 \qq}{H} \epsilon_{i}^{~jk} \pp_j \pd_{\RR^k} H + \left[ \frac{\pp^j \pp_j}{2 H^2} + \frac{2 C^2}{H^2} - 2 \qq^2 \right] \pd_{\RR^i} H ~,
\end{equation}
where we used \eqref{selfdual}.  We also note that the Hamiltonian expressed in the new variables is
\begin{equation}\label{newHam}
H_{\rm c.c.} = m_X \left\{ \half H^{-1} \pp_i \pp^i + 2 \qq \left( H \qq - \frac{\theta_{\rm ym}}{2\pi}\right) + 2 C^2 H^{-1} \right\} ~.
\end{equation}

As we can see from Newton's equation, \eqref{Newton}, there are four different types of forces at play.
\begin{itemize}
\item The first term on the right-hand side of \eqref{Newton} has the typical form of the magnetic force on an electrically charged particle.  The magnetic field is the sum of monopole fields created by the defects and is given by the gradient of the harmonic function, $H$.  The force is proportional to the electric charge $\qq$ of the smooth monopole and vanishes if the smooth monopole carries no electric charge.  Taking into account that the momentum, $\pp_i$ contains a factor of $H$, we see that this force falls off as inverse distance-squared from the defect.  As the smooth monopole approaches the $J^{\rm th}$ defect, it acts to push the monopole in a direction transverse to the plane containing the relative position vector $\vec{R}_J$ and the smooth monopole's instantaneous velocity.  
\item  The second term is a force due to the position-dependent effective mass of the smooth monopole, $m \sim H$.  It is most noticeable when $\qq = 0$ such that the smooth monopole is not prevented from reaching the defects by the electric charge potential barrier.  As the smooth monopole approaches a defect its inertia increases and becomes infinite at $\vec{R}_J = 0$.  Conservation of $H_{\rm c.c.}$ dictates that the speed of the monopole must vanish at this point, and hence it is a turning point of the motion.  
\item The third and fourth terms on the right-hand side of \eqref{Newton}, proportional to $C^2$ and $\qq^2$ respectively, provide competing attractive and repulsive forces on the smooth monopole by each defect.  The attractive force is mediated by the secondary Higgs field while the repulsive force is due to the electrical self-energy of the smooth monopole and originates from a coupling to the long range component of the Higgs field $X$ \cite{Gibbons:1995yw}.  We can also view this term on the same footing at the $\pp^i \pp_i$ term since $\qq$ is the momentum along the circle fiber of Taub--NUT.  
\end{itemize}  

\begin{figure}[th!]
\begin{center}
\includegraphics[width=13cm]{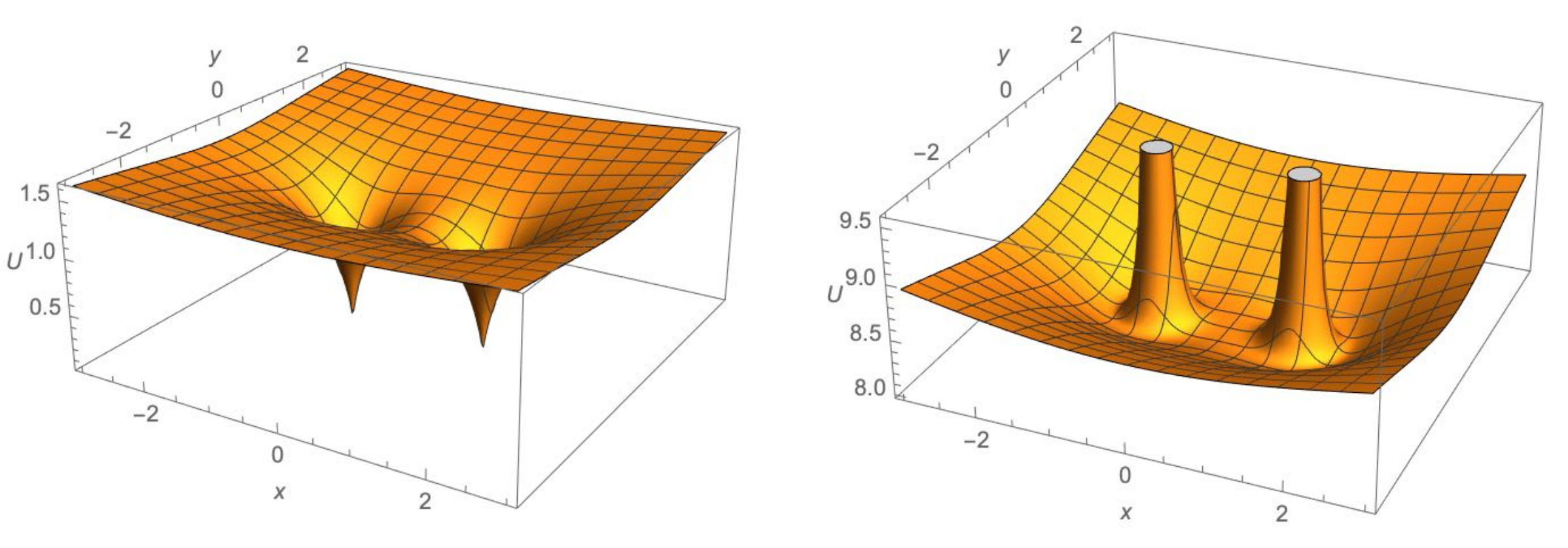}
\end{center}
\caption{The potential energy function \eqref{poten} in the plane $z = 0$ with defects at $\vec{\nu}_1 = (-1,0,0)$ and $\vec{\nu}_2 = (1,0,0)$.  (Here and in the following we use $(x,y,z)$ to denote the Cartesian components of $\vec{\RR}$.)  The left plot has $C = 1$ and $\qq = 0$.  The right plot has $C = 2$ and $\qq = 1$.  If $|\qq| > |C|$ there is no trough in which to contain bound motion.}
\label{fig:poten}
\end{figure}

In Figure \ref{fig:poten} we plot the value of the potential energy function, 
\begin{equation}\label{poten}
U := 4\qq^2 H + 4 C^2 H^{-1}  ~,
\end{equation}
that appears in the Hamiltonian \eqref{newHam} on a two-dimensional plane containing two defects.  The first plot shows the potential energy with $\qq = 0$ and the second plot shows the potential energy with $\qq \neq 0$.  Nonzero electric charge gives rise to a potential barrier that prevents the smooth monopole from passing over a defect.  The function $H^{-2}$ is a nonnegative bounded function on $\mathbbm{R}^3$ that increases to the limiting value $1$ along any ray to infinity.  Hence it follows from \eqref{Newton} that bound motion can exist if and only if $|C| > |\qq|$.  Furthermore, if $\qq = 0$, nontrivial bound motion---\ie other than the static solution $\vec{R} =$ constant---requires $|C| > 0$.  

Equations \eqref{newmomentum} and \eqref{Newton} can be numerically integrated for the smooth monopole's trajectory once the initial position and velocity are specified.  We use the trajectory together with the energy density plots described earlier to construct simulations of the smooth monopole interacting with defects.  Some examples are described in the next subsection.

With only a single defect, additional symmetries enable one to integrate the equations analytically.  In fact, the same set of equations was studied in a different context  in \cite{Jante:2015xra}, where it was shown that the general trajectory is a conic section.  We review and extend this analysis in section \ref{sec:Analytic}.  These analytic results inform the discussion of the various forces above.  
  
\subsection{Simulations}\label{ssec:sim}

We have written several pieces of code in Mathematica for constructing simulations of monopole motion in the presence of defects.  They can be found in the Mathematica notebook included with the ancillary files in this submission.  Brief descriptions of the code and an illustrated example are included in that notebook.  Additionally, the ancillary files contain four high resolution simulations.  Two of these simulations depict bound motion in the presence of a three-defect system, one without electric charge and one with electric charge, and are described here.  The other two movies are a scattering simulation and an oscillating simulation that displays complete screening of the defect.  They are described in subsections \ref{ssec:scattering} and \ref{ssec:timedep} respectively.

We integrate the equations of motion, \eqref{newmomentum} and \eqref{Newton}, numerically to determine the smooth monopole's position and momentum as a function of time.  The inputs are the initial conditions $\vec{\RR}_0,\vec{\pp}_0$ at $\tau = 0$, a set of defect positions $\{\vec{\tilde{\nu}}_J\}$, values for the electric charge $\qq$ and the coupling constant $C$, and the final time $\tau_{\rm max}$ to integrate to.  The resulting trajectory $\vec{\RR}(\tau)$ can be plotted or fed into the code used to produce the energy density plots in Figure \ref{fig:EnergyPlots}.  The code producing the frames for the animation outputs a table of energy density plots.  This table can then be exported as a~.mov or~.avi file using Mathematica's Export command.  The examples included in the ancillary files are based on a sampling rate of 12 frames per unit of time.

\begin{figure}[th!]
\begin{center}
\includegraphics[width=7cm]{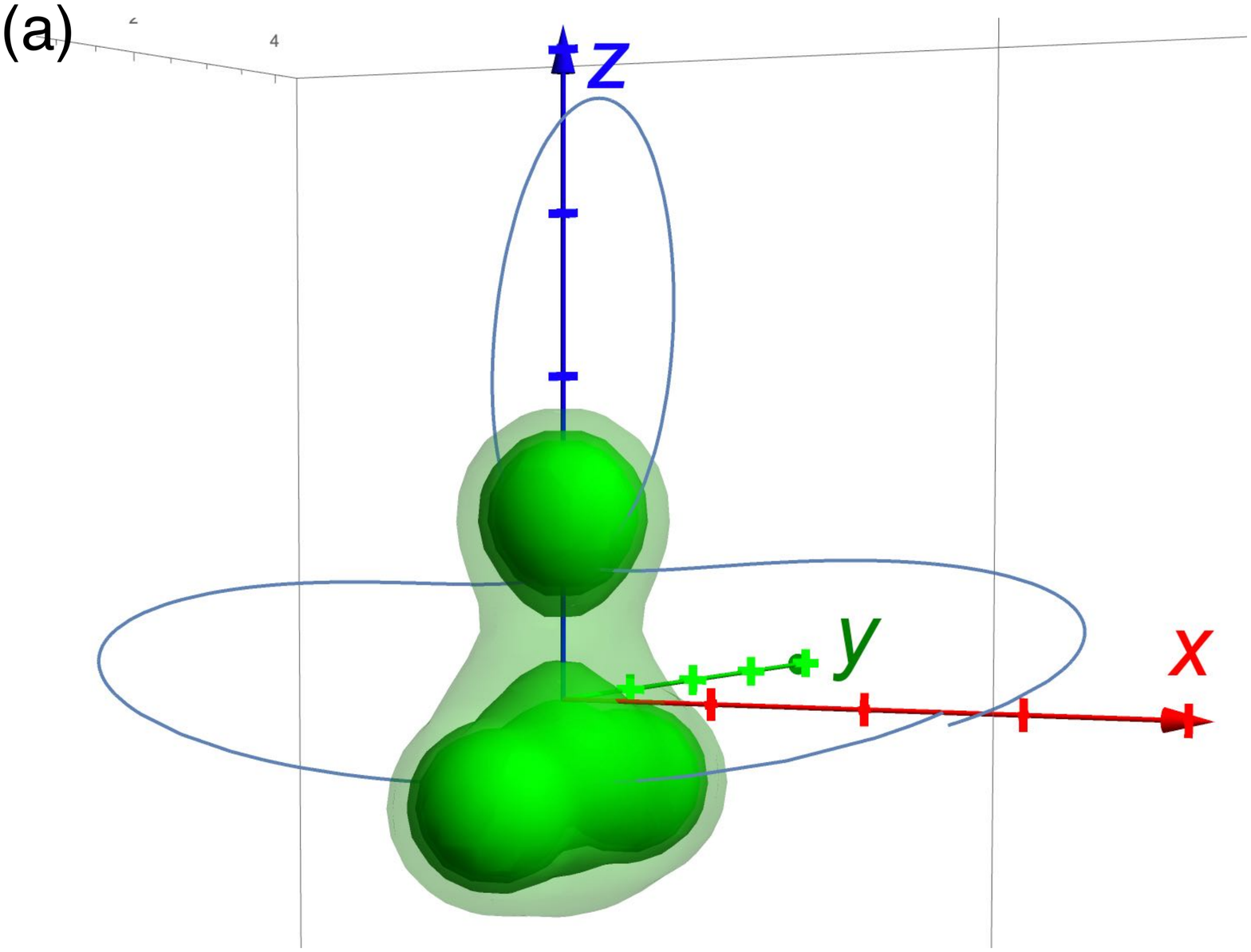} ~ \includegraphics[width=7cm]{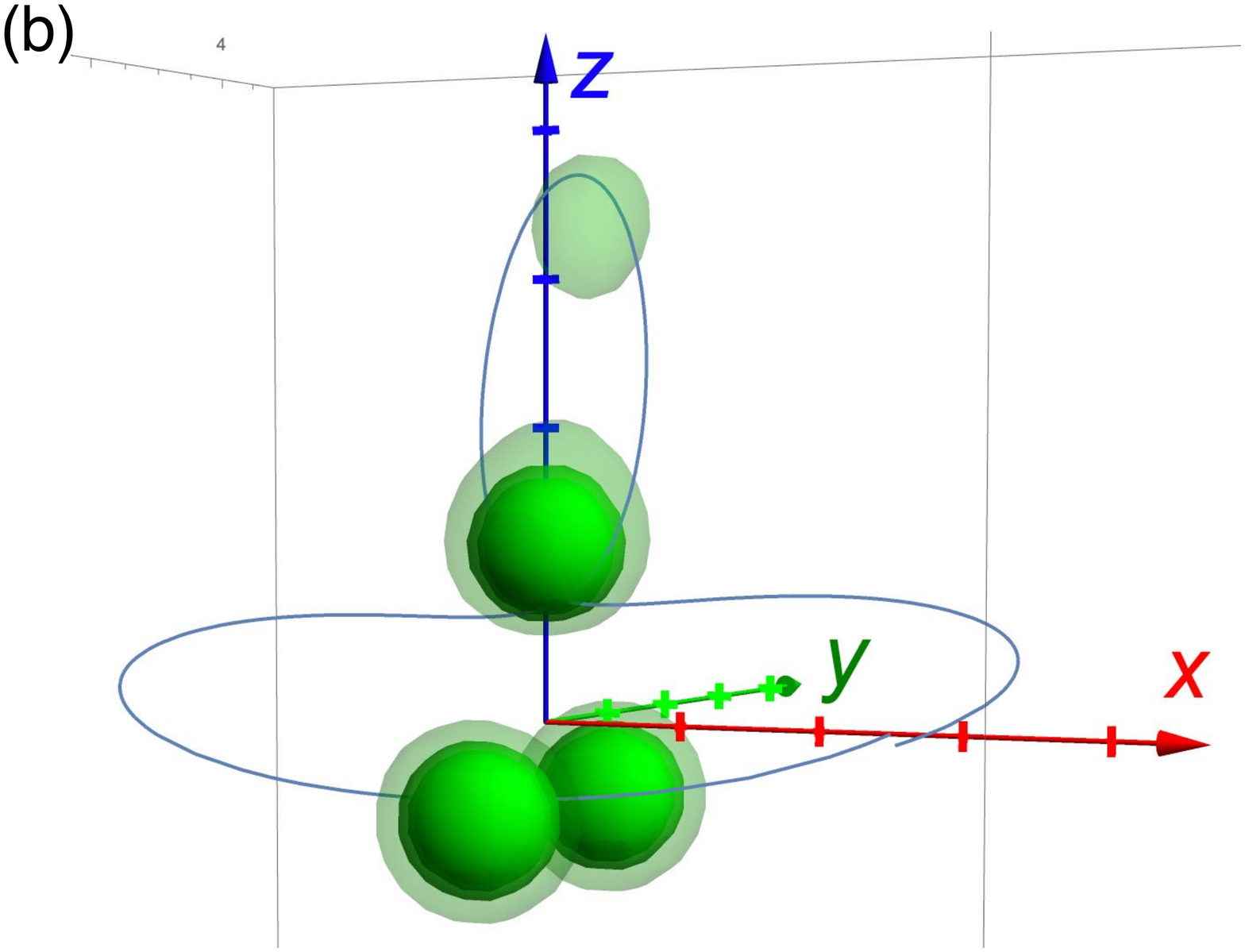}
\end{center}
\caption{Two frames from a simulation with $\qq = 0$ and $C = 3$.  The initial position of the smooth monopole is taken as $\vec{\RR}_0 = (2.5,0,0)$, and the initial momentum is $\vec{\pp}_0 = (-1.1,0,-0.5)$.  The trajectory is shown in blue from $\tau = 0$ to $\tau_{\rm max} = 17.5$, and the frames correspond to $\tau = 1.5$ and $\tau = 10$ for (a) and (b) respectively.}
\label{fig:sim1frames}
\end{figure}

\begin{figure}[th!]
\begin{center}
\includegraphics[width=14.2cm]{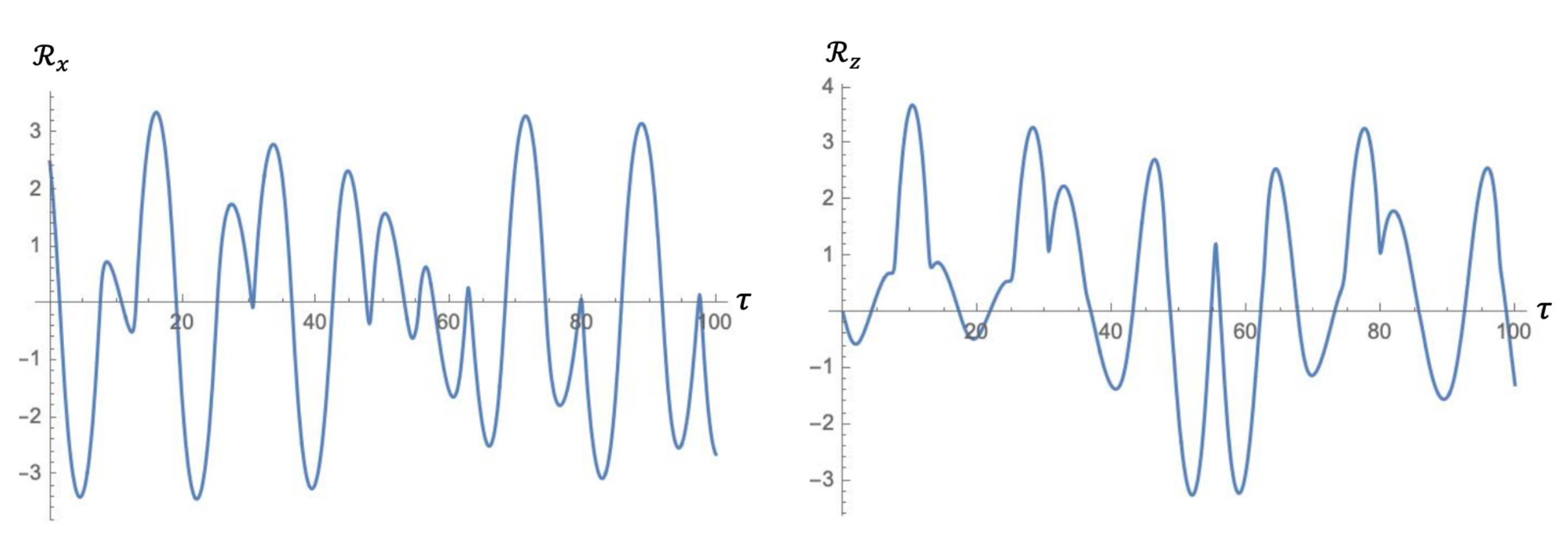}
\end{center}
\caption{The motion in the $\RR_x$-$\RR_z$ plane over longer time scales.}
\label{fig:trajectory}
\end{figure}

In Figure \ref{fig:sim1frames} we show two frames of an animation in the three-defect system with $\qq = 0$ and $C = 3$.  The initial position of the smooth monopole is as in the top of Figure \ref{fig:EnergyPlots}, and the initial velocity is directed back towards the center of the lower two defects.  The trajectory the monopole follows is shown as well.  The motion takes place in the $x$-$z$ plane.  Although the motion is bound, it is not periodic, as the plots in Figure \ref{fig:trajectory} show.

In the case of bound motion such as this, the moduli space trajectory depicted here cannot remain an accurate approximation to the true field theory dynamics for arbitrarily long times.  The reason is that the motion involves continual acceleration which, in the full field theory, will lead to energy loss through radiation.  This energy loss is not captured in the classical truncation to the collective coordinate mechanics that we have employed.  As we discussed in subsection \ref{ssec:ccapprox}, the moduli space trajectory will remain $O(\epsilon)$ close to the true trajectory for times $t \leq T \sim O(1/\epsilon)$ in units of the Higgs vev.  The small parameter $\epsilon$ controls the time variation of the collective coordinates and is naturally identified with $g_{\rm ym}$ in the semiclassical analysis of the quantum Yang--Mills--Higgs theory.  

In terms of the dimensionless time $\tau = \frac{g_{\rm ym}}{2\sqrt{\pi}} m_X t$ introduced in \eqref{dimless}, this result translates to range of times $\tau \leq O(1)$, and one might worry whether the trajectory in Figure \ref{fig:trajectory} can be trusted over one approximate cycle, much less the full range indicated.  This would indeed be an issue if we wished to consider the theory at a small but fixed value of $g_{\rm ym}$.  However, $g_{\rm ym}$ can be chosen arbitrarily small, and nothing we have done so far fixes its value.  Thus, as long as we consider a fixed range of times $\tau \in [0,\tau_{\rm max}]$, where $\tau_{\rm max}$ does not scale with $g_{\rm ym}$ as $g_{\rm ym} \to 0$, then $T = \frac{2\sqrt{\pi}}{g_{\rm ym} m_X} \tau_{\rm max}$ is $O(g_{\rm ym}^{-1})$.  In this way, we may view a $\tau_{\rm max}$ of 20 or 100, as in Figure \ref{fig:trajectory}, as reasonable.  

Note that by sending $g_{\rm ym} \to 0$ we are sending the collective coordinate velocities to zero, via \eqref{gscaling}.  In this language, the observation of the previous paragraph can be phrased as follows.  The effects of radiation on the trajectory over any fixed length of \emph{rescled} time, $\tau \in [0,\tau_{\rm max}]$, can be made arbitrarily small, so long as we are willing to consider arbitrarily slowly moving monopoles with respect to the \emph{true} coordinate time, $t$.  The slowness of these monopoles does not affect the appearance of Figure \ref{fig:trajectory} or the animations, since they are computed with respect to the rescaled time, $\tau$.  From this point of view, one should regard the animations as highly sped-up versions of the ``true'' motion.

\begin{figure}[th!]
\begin{center}
\includegraphics[width=7cm]{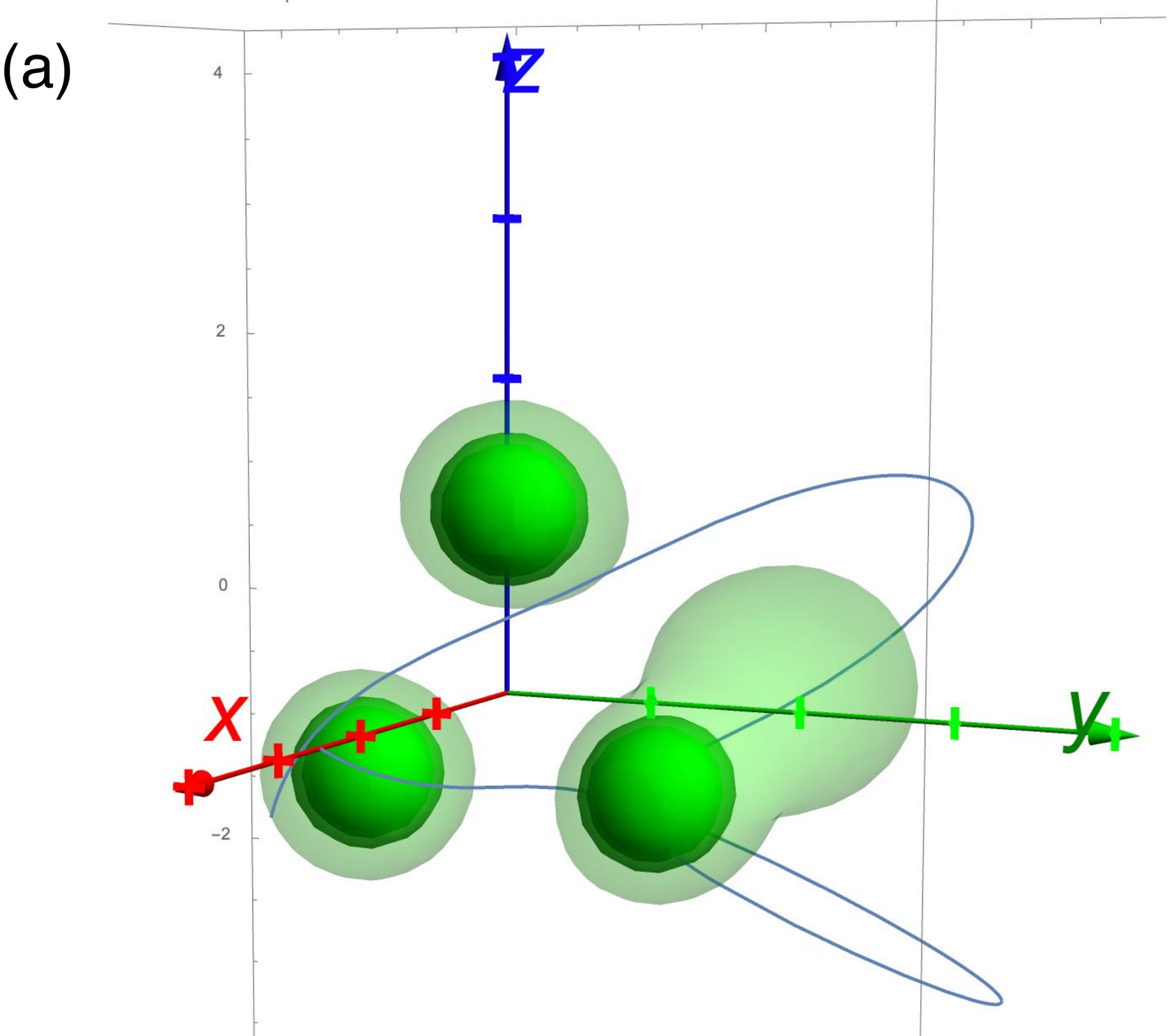} ~ \includegraphics[width=7cm]{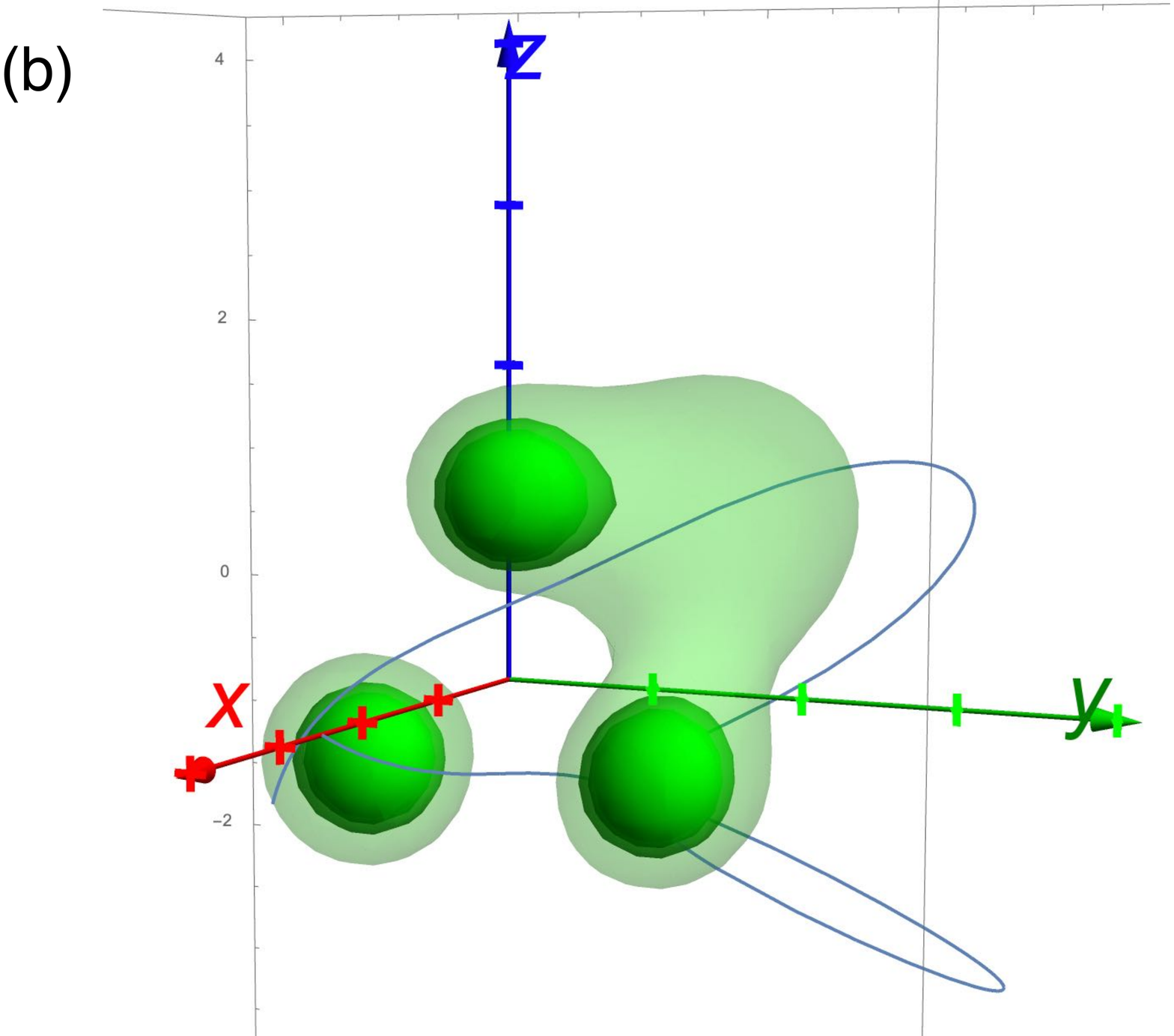}
\end{center}
\caption{Two frames from a simulation with $\qq = 1$ and $C = 3$.  The initial conditions are the same as in the previous example, and the trajectory has been evolved for the same $\tau_{\rm max} = 17.5$.  The frames correspond to $\tau = 10$ and $\tau = 15.2$ for (a) and (b) respectively.}
\label{fig:sim2frames}
\end{figure}

In Figure \ref{fig:sim2frames} we show two frames of a simulation in which the monopole carries electric charge $\qq = 1$.  All other parameters and initial conditions are chosen to be the same as in the first simulation.  The additional magnetic force on the charged monopole causes it to veer outward along the $y$ direction.  The value of $C$ is large enough and the energy low enough, however, that the monopole is drawn back towards the defects and remains bound to them.

We use a different algorithm to determine the level sets of the energy density that will be plotted for the animations versus the individual configurations like those shown in Figure \ref{fig:EnergyPlots}.  This generally results in the smooth monopole being rendered as a single semi-transparent surface in the majority of frames of the animation.  The reason is the following.  In order for the animation to be an accurate representation, the level set values that are used to plot the energy density surfaces should remain fixed from frame to frame.  As the smooth monopole moves away from the defects the overall value of the energy density in its core decreases.  The algorithm ensures that the lowest value for the energy density in the core is always below the lowest level set so that the monopole never disappears.  The remaining level sets increase in regular steps such that the highest one is just below the greatest energy density value that occurs in the smooth monopole's core over the duration of the simulation.\footnote{The code is slightly more sophisticated than this.  When the smooth monopole's position modulus passes very close to a defect the energy value in its ``core'' will blow up.  The code ensures that such excessively large values are not used in the determination of what level surfaces to draw.}  Therefore, multiple surfaces in the smooth monopole tend to only be evident when it is near a defect.  

Each frame in these simulations takes a little over ten minutes to render on a 2018 MacBook Pro with a 2.2 GHz processor and 16 GB of RAM.  Thus, at 12 frames per time unit for 17.5 units, each of these animations took about 1.5 days to finish.

\section{Analytic Results for a Single Defect}\label{sec:Analytic}

Consider the system described by the Hamiltonian \eqref{newHam} with equations of motion \eqref{newmomentum} and \eqref{Newton}.  The electric charge $\qq$ is a constant of motion, as is the total energy, which we write in terms of the dimensionless quantity
\begin{align}\label{ccenergy}
E := m_{X}^{-1} H_{\rm c.c.} + \frac{\qq \theta_{\rm ym}}{\pi} =&~ \half H^{-1} \pp_i \pp^i + 2 \qq^2 H + 2 C^2 H^{-1}  \cr
=&~ \half H \pd_\tau {\vec{\RR}} \cdot \pd_\tau \vec{\RR} +  2 \qq^2 H + 2 C^2 H^{-1}~.
\end{align}
Observe that this energy can be written in either of the forms
\begin{equation}
E = \half H^{-1} \pp_i \pp^i + 2 \left( \qq H^{1/2} \pm C H^{-1/2} \right)^2  \mp 4 \qq C~,
\end{equation}
which implies the lower bound
\begin{equation}\label{Ebound}
E \geq 4 |\qq C|~.
\end{equation}
The bound can only be saturated if the smooth monopole is stationary.

When only a single defect present, such that the harmonic function takes the form
\begin{equation}
H =1 +  \frac{k_{\rm t}}{2 \RR} ~,
\end{equation}
there are additional conserved quantities.  Here we have used translation invariance to place the defect at $\vec{\nu} = 0$, and we also allow for the possibility of an arbitrary defect charge, $|p| = k_{\rm t}$.  This system is equivalent to a model for dyon interactions first considered by Zwanziger \cite{Zwanziger:1969by} and has been encountered in the context of ordinary monopole moduli space dynamics \cite{Lee:2000rp}.  Reference \cite{Jante:2015xra} also provides a recent and detailed treatment.  The additional constants of motion are the angular momentum vector
\begin{equation}\label{angmom}
\vec{J} = \vec{\RR} \times \vec{\pp} + k_{\rm t} \qq \hat{\RR} ~,
\end{equation}
and the Runge-Lenz vector
\begin{equation}\label{RungeLenz}
\vec{K} = \vec{\pp} \times \vec{J} - \frac{k_{\rm t}}{2} (E - 4\qq^2) \hat{R}~.
\end{equation}
Here $\times$ denotes the usual cross product for Euclidean three-vectors, and $\hat{\RR} = \vec{\RR}/\RR$ is the unit vector in the direction of $\vec{\RR}$.

The angular momentum vector receives contributions from the motion of the smooth monopole and from the angular momentum in the electromagnetic field.  The strength of the latter is equal to the Dirac--Schwinger--Zwanziger pairing of the electric and magnetic charges of the monopole and defect.

From the field theory perspective, the origin of the conserved angular momentum is the fact that rotations of the field configuration about the defect map a solution of the Bogomolny equation satisfying all boundary conditions to a new solution, and thus generate a corresponding set of rotational isometries on the moduli space.  The origin of the Runge--Lenz symmetry is due to the extended supersymmetry inherited by the collective coordinate dynamics when the fermions are included in the analysis.  However these additional symmetries can also be understood purely from the point of view of the Hamiltonian particle mechanics, where they are realized as symmetries of the associated six-dimensional phase space.  This point of view is explained nicely in \cite{Jante:2015xra}, where a relationship to the Kepler problem and its Runge--Lenz vector is also discussed.  We refer the reader there for further details.

\subsection{Trajectories}

References \cite{Lee:2000rp,Jante:2015xra} showed how the conserved charges lead to a determination of the trajectories as conic sections.  We review their analysis in this subsection for completeness and since our conventions are slightly different.

First, observe from \eqref{angmom} that
\begin{equation}\label{Jcone}
\vec{J} \cdot \hat{\RR} = k_{\rm t} \qq
\end{equation}
is a constant.  This implies that motion takes place on a cone.  The axis of the cone is $\sgn(\qq) \vec{J}$ and its opening angle, $\theta$, is given by
\begin{equation}\label{costheta}
\cos{\theta} = \frac{\sgn(\qq) \vec{J}\cdot \hat{\RR}}{J} =  \frac{k_{\rm t} |\qq|}{J} =  \frac{k_{\rm t} |\qq|}{\sqrt{ (k_{\rm t} \qq)^2 + |\vec{\RR} \times \vec{\pp}|^2}} ~.
\end{equation}
If $\qq > 0$ then $\vec{J}$ is along the axis of the cone, if $\qq < 0$ then $-\vec{J}$ is along the axis of the cone, and if $\qq = 0$ the ``cone'' is the plane orthogonal to $\vec{J}$.  Since $\qq$ and $J$ are conserved, so is the magnitude $|\vec{\RR} \times \vec{\pp}|$.  However the direction of $\vec{\RR} \times \vec{\pp}$ will in general change along the trajectory.

Next, consider the consequences of the conserved quantity $\vec{K}$.  One finds that
\begin{equation}
\vec{\RR} \cdot \vec{K} = J^2 - (k_{\rm t} \qq)^2 - \frac{k_{\rm t}}{2} (E - 4 \qq^2) \RR ~.
\end{equation}
Since $\vec{\RR} \cdot \vec{J} = \RR k_{\rm t} \qq$ it follows that
\begin{equation}
\left[ k_{\rm t}\qq \vec{K} +  \frac{k_{\rm t}}{2} (E - 4 \qq^2) \vec{J} \right] \cdot \vec{\RR} = k_{\rm t} \qq (J^2 - (k_{\rm t} \qq)^2 )~.
\end{equation}
Hence, defining the conserved vector,
\begin{equation}
\vec{N} := k_{\rm t}\qq \vec{K} +  \frac{k_{\rm t}}{2} (E - 4 \qq^2) \vec{J} ~,
\end{equation}
we have that
\begin{equation}\label{Nplane}
\vec{N} \cdot \vec{\RR} = k_{\rm t} \qq (J^2 - (k_{\rm t} \qq)^2 ) = k_{\rm t} \qq | \vec{\RR} \times \vec{\pp} |^2 ~.
\end{equation}
This is the equation for a plane with outward normal vector $\sgn(\qq) \vec{N}$ and distance to the origin $k_{\rm t} |\qq|  (J^2 - (k_{\rm t} \qq)^2)/N$.  We will compute the magnitude $N$ as well as the angle between $\vec{J}$ and $\vec{N}$, but first we discuss some degenerate cases.

\subsubsection{Motion Along a Ray}

If $\vec{\RR} \times \vec{\pp} = 0$ then the motion takes place along a fixed ray.  Conservation of $\vec{K} \propto \hat{R}$ implies that the smooth monopole cannot pass through the defect or else $\hat{R}$ would flip sign.  As we discussed under \eqref{poten}, bound orbits can only exist if $|C| > |\qq|$.  Since the asymptotic value of the potential energy is $U \to 4 (\qq^2 + C^2)$, we see that the one-dimensional motion will be bounded, and hence oscillatory, when additionally $E < 4 (\qq^2 + C^2)$.  If $E \geq 4 (\qq^2 + C^2)$ the motion will have a single turning point.  When $\qq = 0$ the (inner) turning point will be at the origin, where the mass of the smooth monopole becomes infinitely heavy.  When $|\qq| > 0$ the turning point will be at some distance away from the origin.  We will find the turning point(s) of the motion below when we analyze the generic case.  These formulae will include the case of motion along a ray as a special limit.

\subsubsection{Motion in a Plane Containing the Defect}

If $\qq = 0$ but $\vec{\RR} \times \vec{\pp} \neq 0$, then $\theta = \pi/2$ and the motion takes place in the plane orthogonal to $\vec{J}$, which contains the defect at $\RR = 0$.  This coincides with the plane defined by $\vec{N}$ since $\vec{N} \propto \vec{J}$ when $\qq = 0$.  The Runge--Lenz vector $\vec{K}$ is however interesting in this case.  One finds that $\vec{K} \cdot \vec{J} = 0$ when $\qq = 0$, and hence $\vec{K}$ lies in the plane of motion.  One finds that
\begin{equation}
\vec{K} \cdot \vec{\RR} = J^2 - \frac{k_{\rm t} E}{2} \RR~,
\end{equation}
and hence, defining $\phi$ as the angle measured counterclockwise from $\vec{K}$, we have
\begin{equation}
K \RR \cos{\phi} = J^2 - \frac{k_{\rm t} E}{2} \RR~.
\end{equation}
This is the equation for a conic section,
\begin{equation}\label{conicsection}
\frac{\alpha}{\RR} = 1 + e \cos{\phi}~,
\end{equation}
with semi-latus rectum and eccentricity given by
\begin{equation}\label{latusqzero}
\alpha = \frac{2 J^2}{k_{\rm t} E} ~, \qquad e = \frac{2 K}{k_{\rm t} E}~, \qquad (\qq = 0)~.
\end{equation}

The magnitude of $K$ can be computed by making use of the energy equation to eliminate $\pp^2$.  This is straightforward when $\qq = 0$ and yields
\begin{equation}
K^2 = 2 (E - 2 C^2) J^2 + \frac{k_{\rm t}^2}{4} E^2~, \qquad (\qq = 0)~.
\end{equation}
This results in the eccintricity
\begin{equation}\label{eqzero}
e = \sqrt{1 + \frac{8 J^2 (E - 2 C^2)}{k_{\rm t}^2 E^2}  } ~, \qquad (\qq = 0)~.
\end{equation}
Hence for $E < 2C^2$ the trajectory is an ellipse, for $E = 2C^2$ a parabola, and for $E > 2C^2$ a hyperbola.  The turning points and time dependence follow from setting $\qq = 0$ in the general case below.

\subsubsection{The General Case}

Suppose now that $\qq \neq 0 \neq \vec{\RR} \times \vec{\pp}$.  In this case the plane defined by \eqref{Nplane} does not pass through the origin, and the motion takes place on the intersection of this plane with the cone defined by \eqref{Jcone}.  Thus the trajectory is still a conic section, but it occurs in a plane that does not contain the defect.  To analyze the trajectory in greater detail we need expressions for the length $N$ and the angle $\delta$ between the outward normal to the plane, $\sgn(\qq) \vec{N}$, and the axis of the cone, $\sgn(\qq) \vec{J}$.  A more tedious computation of $K^2$ when $\qq \neq 0$ gives
\begin{equation}
K^2 = 2 (J^2 - (k_{\rm t} \qq)^2) (E - 2 \qq^2 - 2 C^2) + \frac{k_{\rm t}^2}{4} (E - 4 \qq^2)^2~,
\end{equation}
which leads to
\begin{equation}
N^2 = \frac{k_{\rm t}^2}{4} (J^2 - (k_{\rm t} \qq)^2) (E^2 - 16 \qq^2 C^2) ~.
\end{equation}
Meanwhile
\begin{equation}
\vec{N} \cdot \vec{J} = \frac{k_{\rm t}}{2} (J^2 - (k_{\rm t} \qq)^2) (E - 4 \qq^2) ~.
\end{equation}
Hence the angle is
\begin{equation}\label{deltaresult}
\cos{\delta} = \sqrt{ 1 - \frac{(k_{\rm t} \qq)^2}{J^2}} \frac{ (E - 4 \qq^2)}{\sqrt{ E^2 - 16 \qq^2 C^2}} = \sin{\theta} \frac{ (E - 4 \qq^2)}{\sqrt{ E^2 - 16 \qq^2 C^2}}~.
\end{equation}
%

\begin{figure}[th!]
\begin{center}
\includegraphics[width=13cm]{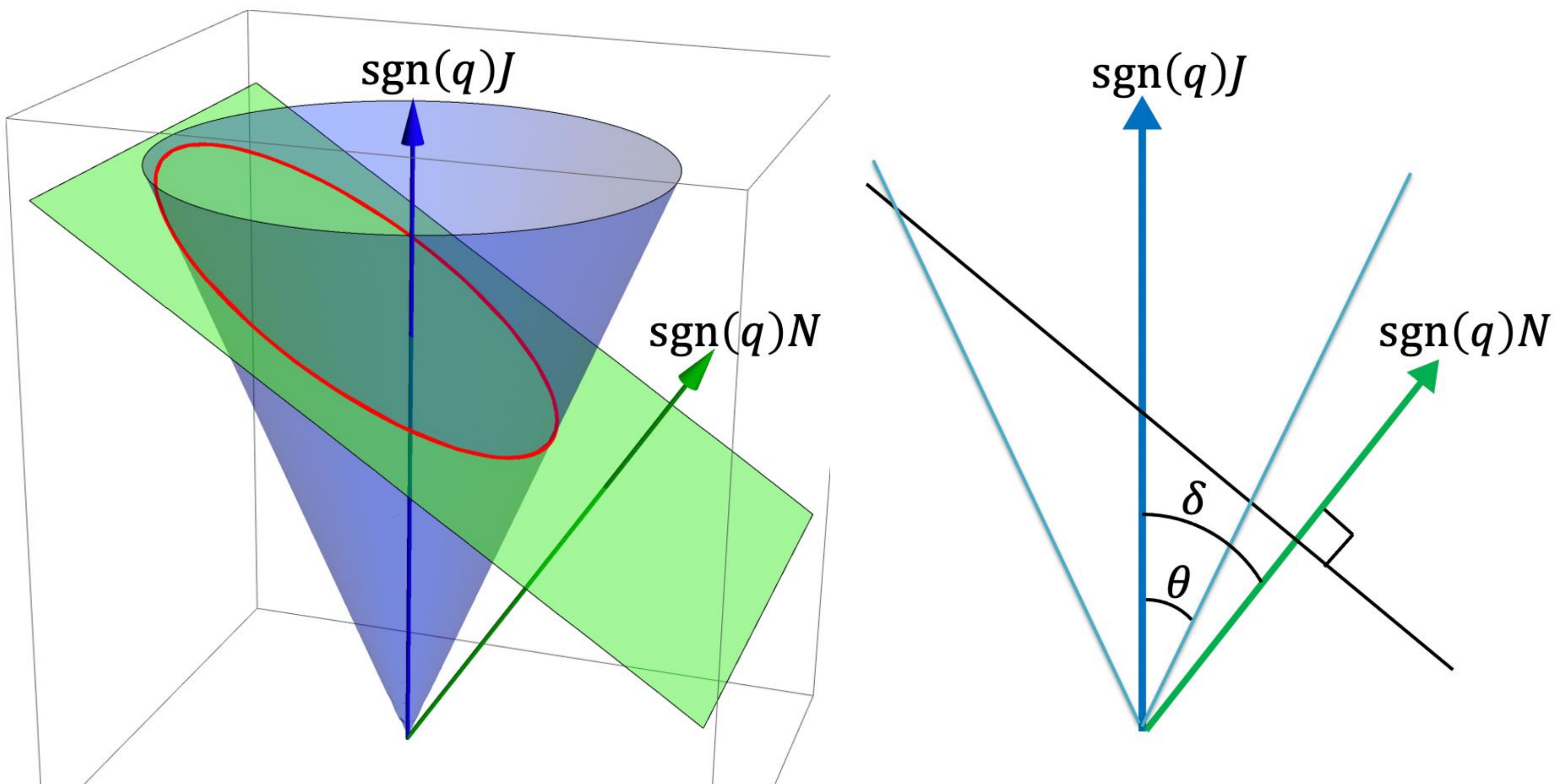}
\end{center}
\caption{The trajectory (highlighted in red) is the intersection of a cone, with axis $\sgn(\qq) \vec{J}$ and opening angle $\theta$, and a plane with outward normal $\sgn(\qq) \vec{N}$ making angle $\delta$ with the axis of the cone.}
\label{fig:conicsection}
\end{figure}

Having obtained $\delta$, we can now determine the conditions for the different types of trajectories.  See Figure \ref{fig:conicsection} for reference.  (The same figure appears in \cite{Jante:2015xra}; we reproduce it here for convenience.)  First, if $\delta \geq \pi/2 + \theta$ there will be no intersection since the plane becomes parallel to the cone at $\delta = \pi/2 + \theta$.  Then the three types of trajectory are
\begin{align}
\frac{\pi}{2} - \theta < \delta < \frac{\pi}{2} + \theta \qquad & \Leftrightarrow \qquad -\sin{\theta} < \cos{\delta} < \sin{\theta}  \qquad  & \Rightarrow \qquad \textrm{hyperbolic}~, \cr
\delta = \frac{\pi}{2} - \theta  \qquad & \Leftrightarrow \qquad  \cos{\delta} = \sin{\theta} \qquad & \Rightarrow \qquad \textrm{parabolic}~, \cr
\delta < \frac{\pi}{2} - \theta \qquad & \Leftrightarrow \qquad \cos{\delta} > \sin{\theta} \qquad & \Rightarrow \qquad \textrm{ellliptic}~.
\end{align}
Note in particular that the parabolic and elliptic cases require $\cos{\delta} > 0$ and therefore, from \eqref{deltaresult}, $E > 4\qq^2$.  The critical values $\delta = \pi/2 \pm \theta$ correspond to $\cos{\delta} = \mp \sin{\theta}$, and therefore by \eqref{deltaresult},
\begin{equation}
\delta = \frac{\pi}{2} \pm \theta \qquad \Leftrightarrow \qquad \frac{ (E_{\rm c} - 4 \qq^2)}{\sqrt{ E_{\rm c}^2 - 16 \qq^2 C^2}} = \mp 1 \qquad \Leftrightarrow \qquad E_{\rm c} = 2 (\qq^2 + C^2)~.
\end{equation}

To determine whether $E$ should be greater or less than $E_{\rm c}$ in each case, consider the functions
\begin{equation}
f_{\pm}(E) = E - 4 \qq^2 \pm \sqrt{ E^2 - 16 \qq^2 C^2} ~,
\end{equation}
which satisfy $f_{\pm}(E_{\rm c}) = 0$.  Then $f_+(E) > 0$ is the condition for solutions to exist while $f_-(E) > 0$ is the condition for bounded motion.  Considering $f_{\pm}'(E)$, we find that $f_+$ is a strictly increasing function while $f_-$ is a strictly decreasing function.

The discussion is divided into two cases.  First suppose that $|C| > |\qq|$.  Then the bound \eqref{Ebound} implies that $E > 4 \qq^2$ and hence that $f_+(E) > 0$.  Thus, \eqref{Ebound} is sufficient to guarantee solutions exist.  It follows from $f_-(E) > 0$ that closed trajectories can only exist for $E < E_{\rm c}$.  In order that this condition be compatible with $E > 4\qq^2$ we require $2 (\qq^2 + C^2) > 4 \qq^2$, which is guaranteed for $|C| > |\qq|$.  Hence we recover the condition discussed under \eqref{poten} for bounded motion to exist.  When $E > E_{\rm c}$, we have that $f_+(E) > 0$ while $f_-(E) < 0$, and therefore the trajectory is hyperbolic.

Now suppose that $|C| < |\qq|$.  In this case $E < E_{\rm c}$ implies $E < 4 \qq^2$.  Thus $f_-$ is already negative and decreases as $E$ increases, so it must be that $f_+$ is the function passing through zero at $E = E_{\rm c}$.  Therefore, in this case, solutions only exist for $E \geq E_{\rm c}$ and are open trajectories.  Hence the results may be summarized as follows:
\begin{align}\label{3cases}
4 |\qq C| < E < 2 (\qq^2 + C^2) \quad \& \quad E > 4\qq^2 \qquad & \Leftrightarrow \qquad \textrm{elliptic}~, \cr
E = 2(\qq^2 + C^2) \qquad & \Leftrightarrow \qquad \textrm{parabolic}~, \cr
E > 2 (\qq^2 + C^2) \qquad & \Leftrightarrow \qquad \textrm{hyperbolic}~.
\end{align}
If $|\qq| > |C|$ the first condition is empty, and elliptic trajectories cannot occur.  These results are consistent with \cite{Jante:2015xra}.

\subsection{Explicit Parameterization, Turning Points, and Time Dependence}\label{ssec:timedep}

Now let us be more explicit about the parameterization of these trajectories.  By rotating our coordinate system, we can always assume that the $z$-axis is in the direction of $\sgn(\qq) \vec{J}$ and that $\vec{N}$ lies in the $x$-$z$ plane.  Then $\theta$, introduced in \eqref{Jcone}, is the polar angle in spherical coordinates, and the motion takes place at fixed $\theta$.  Writing $\vec{\RR}$ in spherical coordinates, the equation for the plane, \eqref{Nplane}, takes the form
\begin{equation}
(N_x \sin{\theta}) \RR \cos{\phi} + (N_z \cos{\theta}) \RR  = k_{\rm t} \qq (J^2 - (k_{\rm t} \qq)^2)~.
\end{equation}
Since $\theta$ is fixed this is an equation relating $\RR$ and $\phi$ which can be brought to a standard form \eqref{conicsection} for a conic section.  We still have the freedom of rotating our coordinate system about the $z$-axis by $180$ degrees to make $N_x$ positive or negative.  We correlate this choice with the sign of $N_z$ so that $N_x/N_z$ is always positive.  Having done so, we then obtain the equation
\begin{equation}\label{conic2}
 \frac{\alpha}{\RR} = 1 + e \cos{\phi} ~,
\end{equation}
with semi-latus rectum and eccentricity given by
\begin{equation}
 \alpha = \frac{k_{\rm t} \qq (J^2 - (k_{\rm t} \qq)^2)}{N_z \cos{\theta}} ~, \qquad e = \tan{\theta} |\tan{\delta}|~.
\end{equation}
Using $N_z = \sgn(\qq) \vec{N} \cdot \vec{J}/J$ and our previous expression for $\cos{\delta}$, \eqref{deltaresult}, we obtain the following:
\begin{equation}\label{newalphae}
\alpha = \frac{2 J^2}{k_{\rm t} (E - 4\qq^2)} ~, \qquad e = \sqrt{1 + \frac{8 J^2 (E - 2 \qq^2 - 2 C^2)}{k_{\rm t}^2 (E - 4\qq^2)^2} } ~.
\end{equation}

These results agree with \eqref{latusqzero} and \eqref{eqzero} when $\qq = 0$.  We also see that the eccentricity is less than one, equal to one, and greater than one in the three cases \eqref{3cases}, confirming the identification of these trajectories.  When $E = 4\qq^2$ both $\alpha$ and $e$ diverge.  In this case \eqref{conic2} reduces to the equation $\RR \cos{\phi} = \alpha/e$ -- \ie the projection of the trajectory onto the $x$-$y$ plane is the straight line given by a constant $x = \alpha/e$.  This is consistent with Figure \ref{fig:conicsection} since this case corresponds to $\delta =\pi/2$.  When $E$ is less than $4\qq^2$ the semi-latus rectum is negative, which indicates that the trajectory bends away from the defect rather than bending around it.  This is again consistent with Figure \ref{fig:conicsection} when $\delta > \pi/2$.

We turn to the energy equation, \eqref{ccenergy}, in order to determine the time dependence.  On the one hand, solving this equation for $(\pd_{\tau} \vec{\RR})^2$ yields
\begin{align}
(\pd_{\tau} \vec{\RR})^2 =&~ \frac{2 (E - 2 \qq^2 H - 2 C^2 H^{-1})}{H}  \cr
=&~  \frac{ 2 (E - 2 \qq^2 - 2 C^2) \RR^2 + k_{\rm t} (E - 4\qq^2) \RR - k_{\rm t}^2 q^2}{ (\RR + \frac{k_{\rm t}}{2})^2} ~.
\end{align}
On the other hand, restricting $(\pd_{\tau} \vec{\RR})^2$ to the trajectory defined by constant $\theta$ and \eqref{conic2} yields
\begin{align}
(\pd_{\tau} \vec{\RR})^2 =&~ (\pd_\tau \RR)^2 + \RR^2 \sin^2{\theta} \left(\frac{d\phi}{d\RR}\right) (\pd_\tau \RR)^2 \cr
=&~ \left( \frac{ (e^2 - 1) \RR^2 + 2\alpha \RR - \alpha^2 \cos^2{\theta}}{(e^2 - 1) \RR^2 + 2\alpha \RR - \alpha^2} \right) (\pd_\tau \RR)^2 ~.
\end{align}
Setting the two equal we obtain an expression for $\pd_{\tau} \RR$.

This expression simplifies thanks to the identity
\begin{equation}
\frac{2 (E - 2 \qq^2 - 2 C^2) \RR^2 + k_{\rm t} (E - 4 \qq^2) \RR - k_{\rm t}^2 \qq^2}{ (e^2 -1) \RR^2 + 2\alpha \RR - \alpha^2 \cos^2{\theta}} = \frac{k_{\rm t}^2 (E - 4\qq^2)^2}{4 J^2} ~,
\end{equation}
which can be demonstrated by using \eqref{newalphae} and \eqref{costheta}.  Hence we find
\begin{equation}
(\pd_\tau \RR)^2 = \frac{k_{\rm t}^2 (E - 4\qq^2)^2}{4 J^2} \frac{ [ (e^2 - 1) \RR^2 + 2\alpha \RR - \alpha^2] }{(\RR + \frac{k_{\rm t}}{2} )^2} ~.
\end{equation}

Now we consider the turning points of the motion where the numerator on the right-hand side is zero.  The roots of the polynomial in square brackets are
\begin{equation}
\RR = \frac{-\alpha \pm |\alpha| e}{e^2 - 1}~, \quad (e \neq 1)~, \qquad \textrm{or} \qquad \RR = \frac{\alpha}{2}~, \quad (e = 1)~.
\end{equation}
In the case of elliptic motion, $e < 1$ and, from our previous discussion, $\alpha$ is necessarily positive.  Hence both roots are positive and real.  These represent the perigee and apogee of the motion.  In the case of parabolic motion, $\alpha$ is positive, and $\RR = \alpha/2$ is the distance of closest approach.  In the case of hyperbolic motion, $e > 1$ and $\alpha$ can have either sign.  One root is positive and the other is negative, but which is which depends on the sign of $\alpha$.  The positive root is the distance of closest approach.  Setting
\begin{equation}\label{turningpoints}
\RR_{\pm} := \frac{|\alpha|}{\sgn(\alpha) \pm e}~,
\end{equation}
the results can be summarized as follows:
\begin{align}
\textrm{elliptic:} & \qquad 0 \leq \RR_- < \RR \leq \RR_+ ~, \cr
\textrm{parabolic:} & \qquad 0 < \frac{\alpha}{2} \leq \RR ~, \cr
\textrm{hyperbolic:} & \qquad \RR_- < 0 < \RR_+ \leq \RR ~.
\end{align}

The time dependence is then determined by
\begin{align}\label{motion1}
\textrm{elliptic:} \qquad & \frac{d\RR (\RR + \frac{k_{\rm t}}{2})}{\sqrt{ (\RR_+ - \RR)(\RR - \RR_-)} }  = \pm \sqrt{2 (2 \qq^2 + 2 C^2 - E)} \, d\tau ~, \cr 
\textrm{parabolic:} \qquad & \frac{d\RR (\RR + \frac{k_{\rm t}}{2})}{ \sqrt{ \RR - \frac{\alpha}{2}}} = \pm \sqrt{2 k_{\rm t} (C^2 - \qq^2)} \, d\tau ~, \cr
\textrm{hyperbolic:} \qquad & \frac{ d\RR  (\RR + \frac{k_{\rm t}}{2})}{ \sqrt{ (\RR - \RR_+)(\RR - \RR_-)}} = \pm \sqrt{2 (E - 2 \qq^2 - 2 C^2)} \, d\tau ~,
\end{align}
where the sign choice corresponds to the outward or inward part of the trajectory respectively, and the required indefinite integrals are
\begin{align}
\mathcal{I}_{\rm e}(\RR)  :=&~  \int \frac{d\RR (\RR + \frac{k_{\rm t}}{2})}{\sqrt{ (\RR_+ - \RR)(\RR - \RR_-)} } \cr
=&~ - \left\{ \sqrt{ (\RR_+ - \RR)(\RR - \RR_-)} + (k_{\rm t} + \RR_+ + \RR_-) \arcsin\left(\sqrt{\frac{\RR_+ - \RR}{\RR_+ - \RR_-}}\right) \right\} ~, \cr 
\mathcal{I}_{\rm p}(\RR) :=&~ \int \frac{d\RR (\RR + \frac{k_{\rm t}}{2})}{ \sqrt{ \RR - \frac{\alpha}{2}}}  = \left(k_{\rm t} + \frac{2}{3}(\RR + \alpha) \right) \sqrt{\RR - \frac{\alpha}{2}} ~,   \cr
\mathcal{I}_{\rm h}(\RR) : =&~ \int \frac{ d\RR  (\RR + \frac{k_{\rm t}}{2})}{ \sqrt{ (\RR - \RR_+)(\RR - \RR_-)}}  \cr
=&~  \sqrt{(\RR - \RR_+)(\RR - \RR_-)} + (k_{\rm t} + \RR_+ + \RR_-) \ln\left( \frac{\sqrt{\RR - \RR_+} + \sqrt{\RR - \RR_-}}{\sqrt{\RR_+-\RR_-}}\right)~. \qquad
\end{align}

As in the Kepler problem, the equation for $\RR$ as a function of $\tau$ is transcendental.  We can, however, obtain an analytical expression for the period of an elliptical orbit:
\begin{align}\label{period}
T =&~ \frac{2}{\sqrt{2 (2 \qq^2 + 2 C^2 - E)}} (\mathcal{I}_{\rm e}(\RR_+) - \mathcal{I}_{\rm e}(\RR_-)) = \frac{\pi}{\sqrt{2 (2 \qq^2 + 2 C^2 - E)}} (k_{\rm t} + \RR_+ + \RR_-) \cr
=&~  \frac{\pi}{\sqrt{2 (2 \qq^2 + 2 C^2 - E)}} \left( k_{\rm t} + \frac{2\alpha}{1-e^2} \right) = \frac{\pi k_{\rm t} (4C^2 - E)}{(2 (2 \qq^2 + 2 C^2 - E))^{3/2}} ~.  \raisetag{20pt}
\end{align}
The equations for the turning points \eqref{turningpoints}, the time dependence \eqref{motion1}, and the period \eqref{period} are valid in all cases, including the case of one-dimensional motion along a ray.  In particular, $T$ gives the oscillation period in this case.

\begin{figure}[th!]
\begin{center}
\includegraphics[width=7cm]{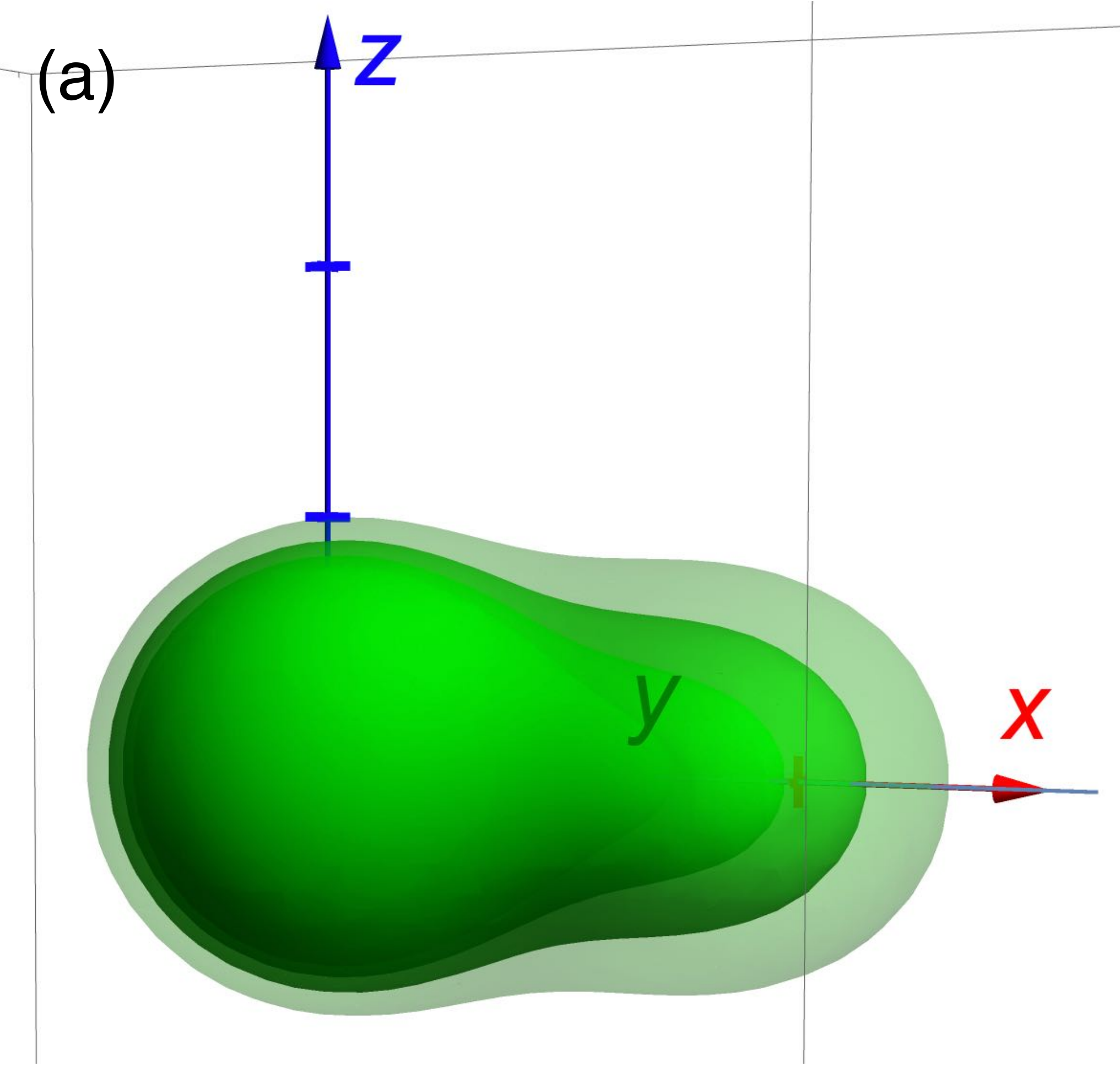} ~ \includegraphics[width=7cm]{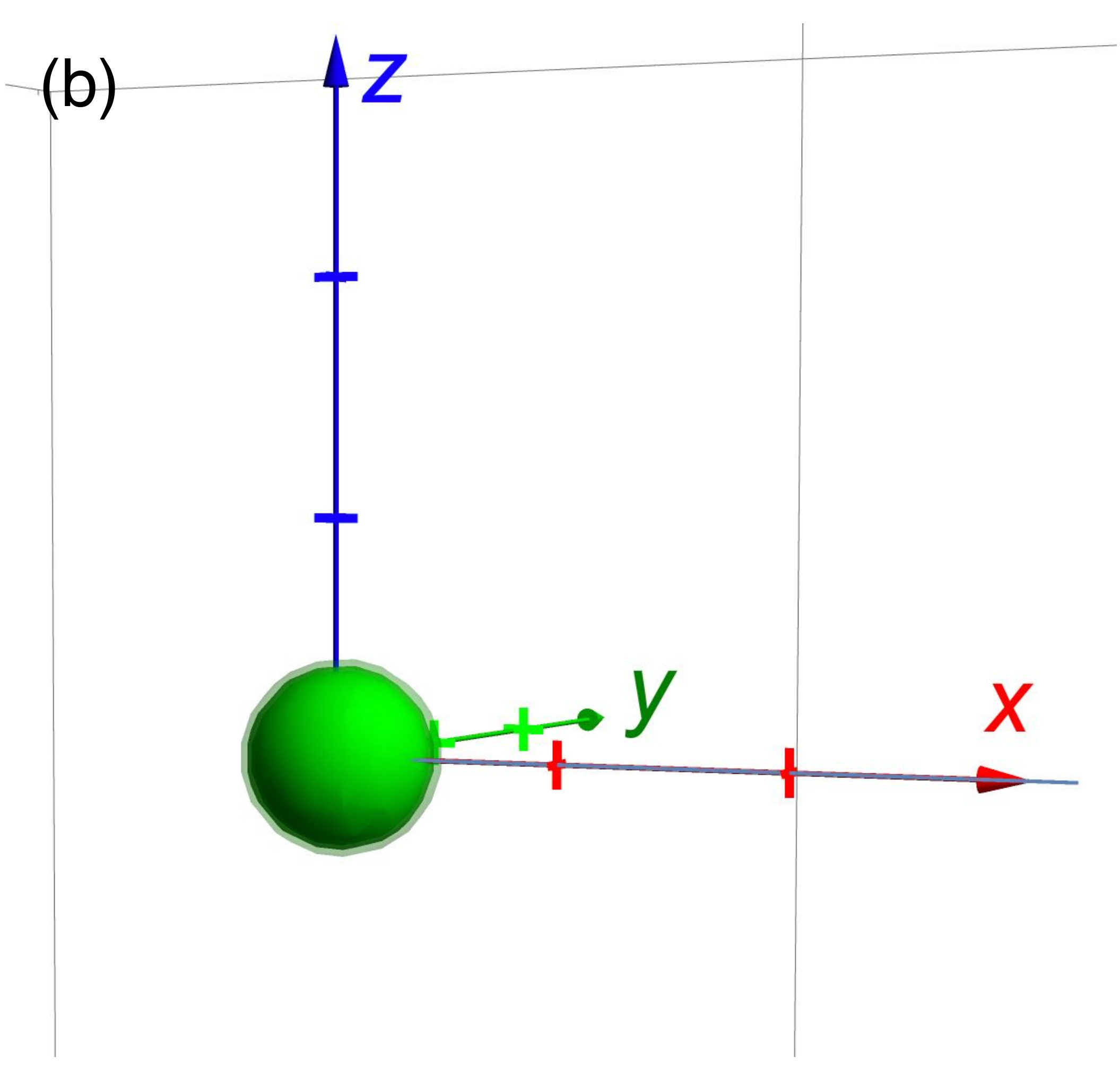} \\  
\includegraphics[width=7cm]{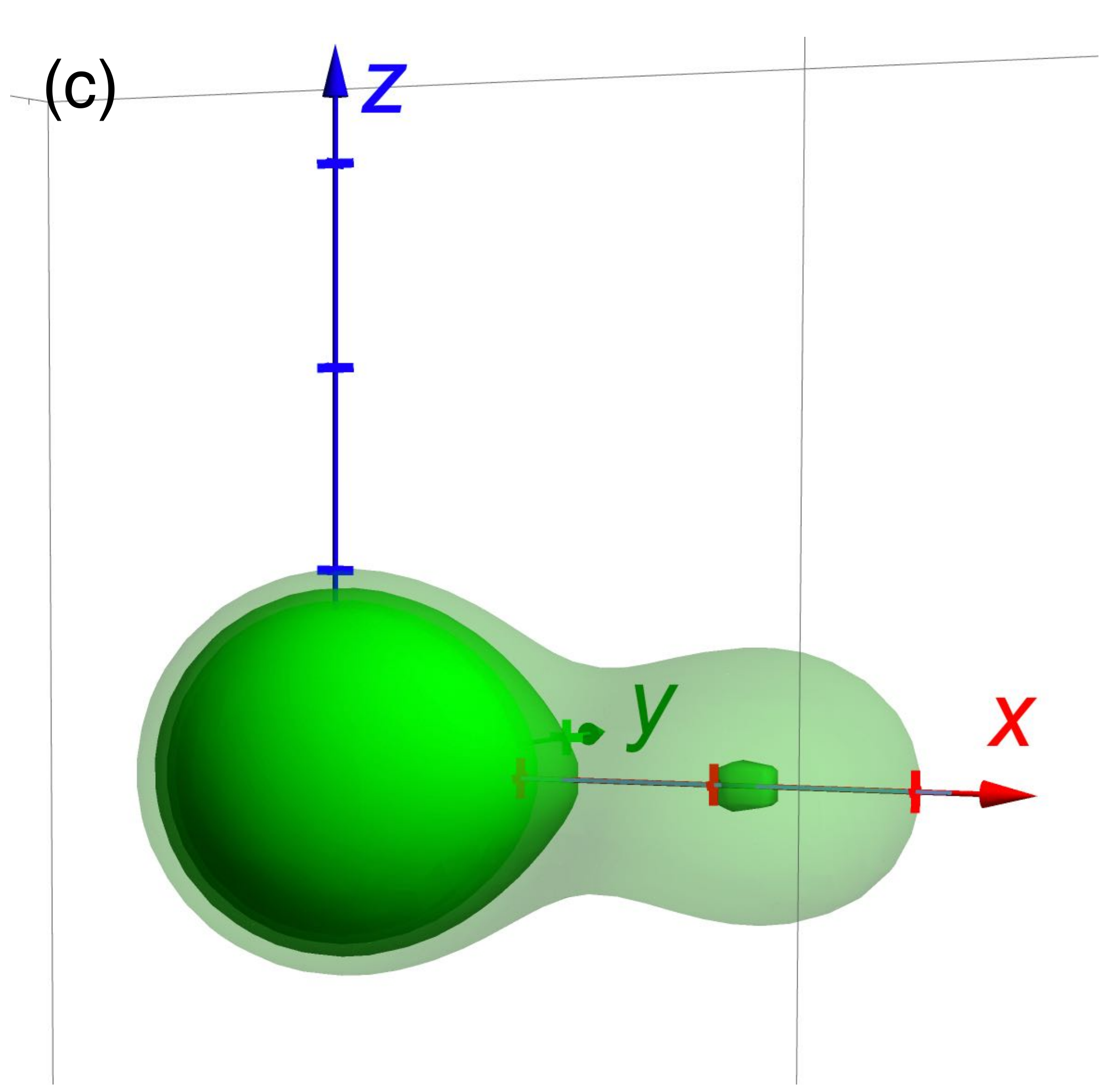} ~ \includegraphics[width=7cm]{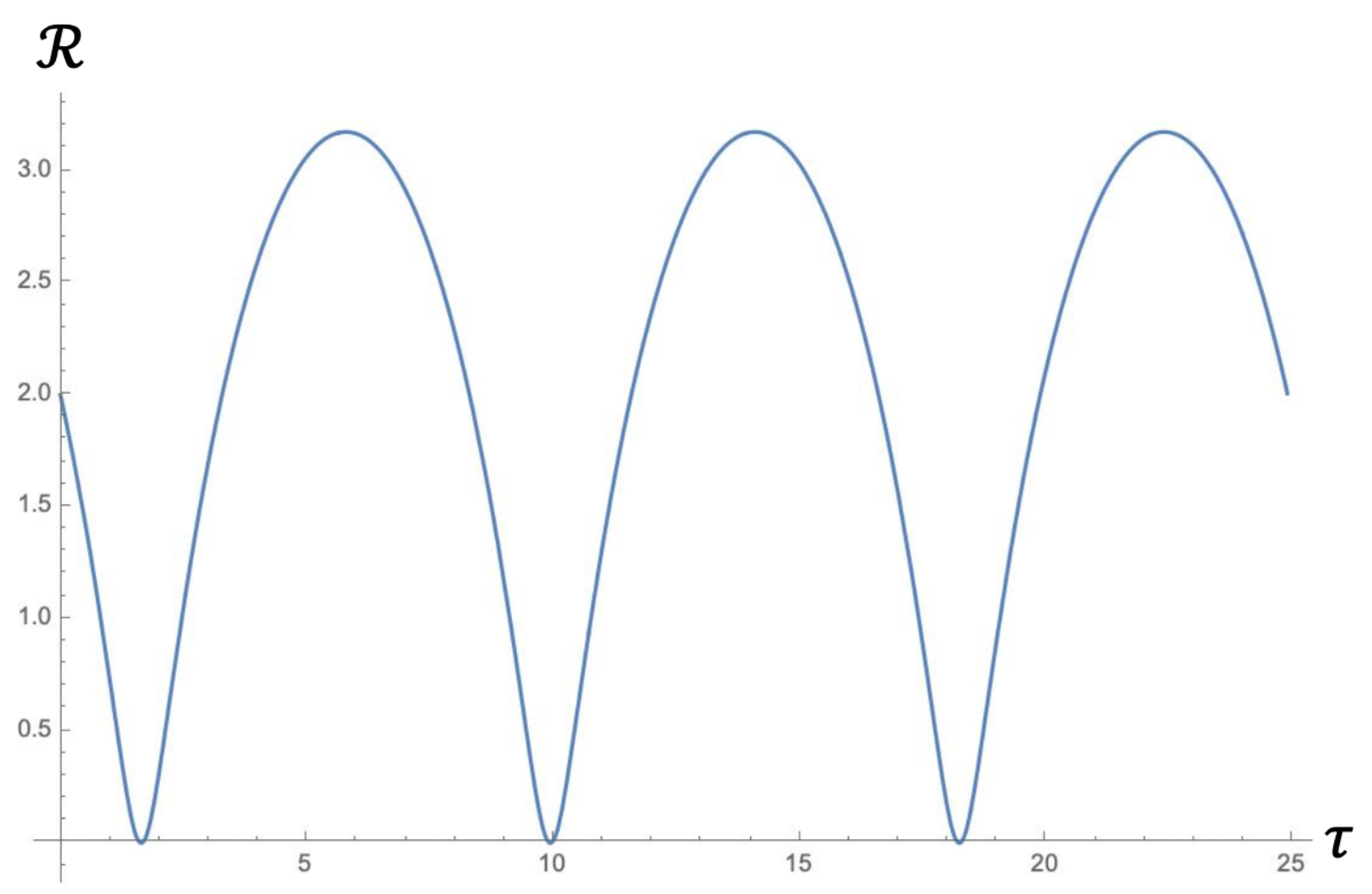}
\end{center}
\caption{Three frames from an oscillating solution and a plot of the defect-monopole distance as a function of time.  The configuration has $k_{\rm t} = 2$, $\qq = 0$, and $C = 2$.  The initial position and momentum are $\vec{\RR}_0 = (2,0,0)$ and $\vec{\pp}_0 = (-1,0,0)$.  This gives $E = 6.08$ and a period of $T = 8.30$, which is consistent with \eqref{period}.  Frame (a) corresponds to $\tau = 0$, frame (b) to $\tau = 1.75$, and frame (c) to $\tau = 3.98$.}
\label{fig:sim3frames}
\end{figure}

The ancillary material includes an animation of an oscillating solution with $k_{\rm t} = 2$, $\qq = 0$, and $C = 2$.  When $\qq = 0$ the inner turning point is at $\RR_- = 0$, atop the defect.  Furthermore, when the defect charge is $k_{\rm t} = 2$, the smooth monopole can completely screen the defect: at $\vec{\RR} = 0$ the solution becomes the trivial one with vanishing energy density.  The vanishing of the asymptotic magnetic charge \eqref{magcharge} when $k_{\rm t} = 2$ is consistent with the possibility of this configuration being a point in the moduli space.  However, the evolution is smooth through this singularity and the monopole and defect reemerge.  In Figure \ref{fig:sim3frames} we show three frames from the animation and a plot of the monopole-defect distance as a function of time.

\subsection{Scattering off the Defect}\label{ssec:scattering}

In this final subsection we analyze the scattering problem for the smooth monopole off of the defect at the origin.  Reference \cite{Lee:2000rp} previously gave the differential scattering cross section for a mathematically equivalent problem by generalizing the results in \cite{Gibbons:1986df} to the case a nonvanishing attractive potential (\ie\ $C \neq 0$ in our language).  We take a different and elementary approach based on the classical trajectory and a rotation between reference frames.

\begin{figure}[th!]
\begin{center}
\includegraphics[width=8cm]{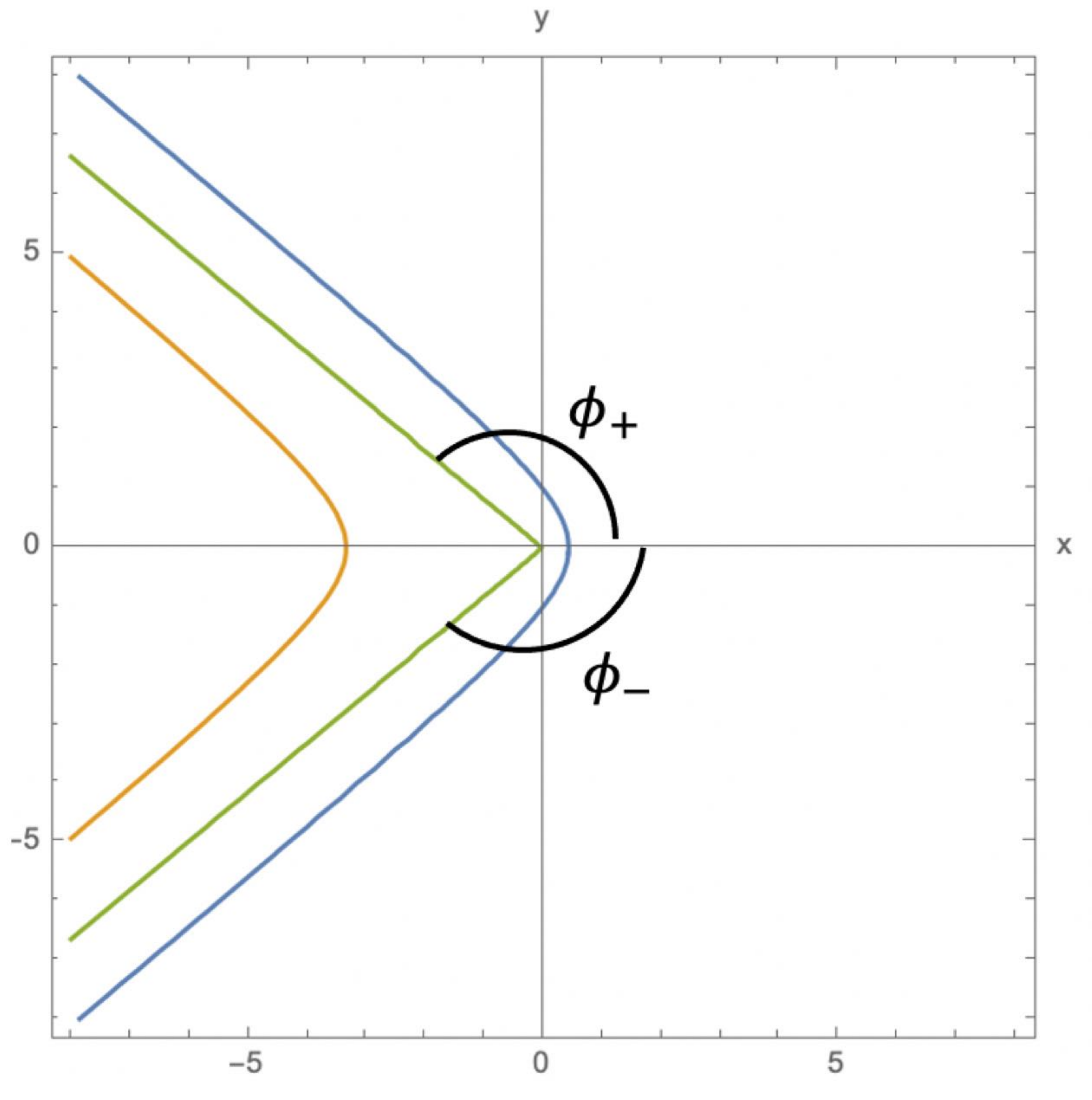}
\end{center}
\caption{Here we plot two conic sections for $e = 1.3$ with $\alpha = +1$ (blue) and $\alpha = -1$ (orange).  The lines defining the angles $\phi_{\pm}$ are also plotted in green.}
\label{fig:scattering}
\end{figure}

Scattering is easy to analyze in the adapted coordinate system of Figure \ref{fig:conicsection}.  Motion takes place at a constant value of the polar angle $\theta$, while the projection of the trajectory into the $x$-$y$ plane is a hyperbola that starts in the second quadrant and ends in the third quadrant or vice versa.  The asymptotic initial and final angles, $\phi_{\pm}$, are the two solutions to 
\begin{equation}
\cos{\phi_{\pm}} = -1/e~,
\end{equation}
in the range $(\pi/2, 3\pi/2)$.  See Figure \ref{fig:scattering}.

Therefore our approach is the following.  We set up some initial data---an incoming velocity and an impact parameter---for the scattering problem in the ``lab'' frame, compute the values of the conserved quantities, $E,\vec{J},\vec{N}$, and rotate the incoming direction to the adapted coordinate system, where it must match up with one of
\begin{equation}\label{npm}
\hat{n}_{\pm} = \left( \begin{array}{c} \cos{\phi_{\pm}} \sin{\theta} \\ \sin{\phi_{\pm}} \sin{\theta} \\ \cos{\theta} \end{array} \right) ~.
\end{equation}
The other of these two is then the outgoing direction in the adapted frame, which we finally rotate back to the lab frame to obtain the outgoing direction in the lab frame.  We will refer to the lab frame as the primed frame in the following.

Without loss of generality, we take our initial data to be
\begin{equation}
\vec{\RR}_{0}' = \left( \begin{array}{c} b \\ 0 \\ +\infty \end{array} \right) ~, \qquad \vec{\pp}_{0}^{\,\prime} = \left( \begin{array}{c} 0 \\ 0 \\ - v \end{array} \right)~,
\end{equation}
with $v > 0$ and $b \geq 0$, so that the smooth monopole is coming in parallel to the $z$-axis, a distance $b$ from it in the direction of the positive $x$-axis.  Using these, we obtain the conserved quantities
\begin{align}
E =&~ \half v^2 + 2 \qq^2 + 2 C^2 ~, \qquad \vec{J}^{\, \prime} = \left( \begin{array}{c} 0 \\ bv \\ k_{\rm t} \qq \end{array} \right) ~, \qquad  \vec{N}^{\prime} = \left( \begin{array}{c} N_{x}' \\ N_{y}' \\ 0 \end{array} \right) \quad \textrm{where} \cr
N_{x}' =&~ k_t \qq b v^2 ~, \qquad N_{y}' = \frac{k_{\rm t}}{2} \left( \tfrac{1}{2} v^2 + 2 C^2 - 2 \qq^2 \right) bv~.
\end{align}

Since $\sgn(\qq) \vec{J}^{\,\prime}$ has no $x$-component, we first rotate clockwise by angle $\theta_0$ about the $x$-axis to line up $\sgn(\qq) \vec{J}^{\, \prime}$ with the $z$-axis.  This determines $\theta_0$ such that
\begin{equation}
\sin{\theta_0} = \frac{\sgn(\qq) b v}{\sqrt{ (bv)^2 + (k_{\rm t} \qq)^2}} ~, \qquad \cos{\theta_0} = \frac{k_{\rm t} |\qq|}{\sqrt{ (bv)^2 + (k_{\rm t} \qq)^2} } ~.
\end{equation}
Note that $\theta_0 = \sgn(\qq) \theta$.  Now we rotate counterclockwise about the new $z$-axis by angle $\phi_0$, such that
\begin{align}
\vec{N} =  R_{\phi_0} R_{\theta_0} \vec{N}^{\prime} =&~ \left( \begin{array}{c c c} \cos{\phi_0} & \sin{\phi_0} & 0 \\ - \sin{\phi_0} & \cos{\phi}_0 & 0 \\ 0 & 0 & 1 \end{array}\right) \left( \begin{array}{c c c} 1 & 0 & 0 \\ 0 & \cos{\theta_0} & -\sin{\theta_0} \\ 0 & \sin{\theta_0} & \cos{\theta_0} \end{array} \right) \left( \begin{array}{c} N_{x}' \\ N_{y}' \\ 0 \end{array} \right) \cr
=&~  \left( \begin{array}{c} \cos{\phi_0} N_{x}' + \sin{\phi_0} \cos{\theta_0} N_{y}' \\ - \sin{\phi_0} N_{x}' + \cos{\phi_0} \cos{\theta_0} N_{y}' \\ \sin{\theta_0} N_{y}' \end{array} \right) ~.
\end{align}
By requiring $N_y = 0$ and $\sgn(N_x) \sgn(N_z) = 1$ we determine
\begin{equation}
\sin{\phi_0} =  \frac{ \sgn(N_x') |N_y'| \cos{\theta_0}}{\sqrt{ N_{x}^{\prime 2} + N_{y}^{\prime 2} \cos^2{\theta_0}}} \qquad \cos{\phi_0} = \frac{ \sgn(N_y') |N_{x}'|}{\sqrt{ N_{x}^{\prime 2} + N_{y}^{\prime 2} \cos^2{\theta_0}}} ~,
\end{equation}
where we note that 
\begin{equation}
\sgn(N_{x}') = \sgn(\qq)~, \qquad \sgn(N_{y}') = \sgn\left( \tfrac{1}{2} v^2 + 2 C^2 - 2 \qq^2\right) = \sgn(E - 4\qq^2)~.
\end{equation}

With the transformation from lab to adapted coordinate system in hand, we can determine the incoming direction, $\hat{n}_{\rm in}' = \hat{k}$ in the adapted coordinate system:
\begin{equation}
\hat{n}_{\rm in} = R_{\phi_0} R_{\theta_0} \hat{n}_{\rm in}' = \left( \begin{array}{c} - \sin{\phi_0} \sin{\theta_0} \\ - \cos{\phi_0} \sin{\theta_0} \\ \cos{\theta_0} \end{array} \right) ~.
\end{equation}
As a check, one can use the above results to verify that
\begin{equation}
(n_{\rm in})_x = -\frac{1}{e} \sin{\theta} ~.
\end{equation}
We find that $\hat{n}_{\rm in} = \hat{n}_\pm$, \eqref{npm}, for $\sgn(N_{x}') \sgn(N_{y}') = \mp 1$ respectively.  Regardless, the outgoing direction in the adapted coordinate system is therefore
\begin{equation}
\hat{n}_{\rm out} =    \left( \begin{array}{c} - \sin{\phi_0} \sin{\theta_0} \\ \cos{\phi_0} \sin{\theta_0} \\ \cos{\theta_0} \end{array} \right) ~.
\end{equation}

Finally we rotate this vector back to the lab frame to determine the outgoing direction of the smooth monopole after the interaction:
\begin{equation}
\hat{n}_{\rm out}' = R_{\phi_0}^{-1} R_{\theta_0}^{-1} \hat{n}_{\rm out} ~.
\end{equation}
We find the following components:
\begin{align}\label{nout}
(n_{\rm out}')_x =&~ \frac{- 4 k_{\rm t} b v^2 (\half v^2 + 2C^2 - 2 \qq^2)}{ k_{\rm t}^2 (\half v^2 + 2C^2 - 2 \qq^2)^2 + 4 v^2 ( (bv)^2 + (k_{\rm t} q)^2)} ~, \nonumber \\[2ex]
(n_{\rm out}')_y =&~ \frac{ 8 k_{\rm t} \qq b v^3}{ k_{\rm t}^2 (\half v^2 + 2C^2 - 2 \qq^2)^2 + 4 v^2 ( (bv)^2 + (k_{\rm t} q)^2) } ~, \nonumber \\[2ex]
(n_{\rm out}')_z =&~ \frac{k_{\rm t}^2  (\half v^2 + 2 C^2 - 2 \qq^2)^2 + 4 v^2 ( (k_t \qq)^2 - (bv)^2)}{  k_{\rm t}^2  (\half v^2 + 2 C^2 - 2 \qq^2)^2 + 4 v^2  ( (bv)^2 + (k_{\rm t} \qq)^2)} ~.
\end{align}
Some comments on the result are:
\begin{itemize}
\item If $\qq = 0$ the y-component vanishes, and this is consistent with the fact that the scattering should take place in the $x$-$z$ plane, which is the plane containing the defect.  The sign of $\qq$ determines whether the monopole scatters above or below the initial plane of motion.
\item If $E - 4\qq^2 = \half v^2 + 2 C^2 - 2 \qq^2$ is positive, then the $x$-component is negative, meaning that the trajectory bends around the defect, while if it is negative then the trajectory bends away from the defect.  Recalling that $E = 4\qq^2$ is the condition that $\delta = \pi/2$, this behavior is consistent with Figure \ref{fig:conicsection} and the comments under \eqref{newalphae}.
\item The sign of the $z$-component can always be made negative by choosing $b$ large enough.  This makes sense: if $b$ is large, then the trajectory shouldn't be much affected by the defect, and hence the direction of the monopole's final velocity should be close to that of the initial velocity.
\item  If $b = 0$ the outward direction simplifies to $\hat{n}_{\rm out}' = \hat{k}$.  The smooth monopole approaches the defect, comes to a stop, and reverses.  This corresponds to the case of motion along a ray, discussed earlier.
\end{itemize}
%

\begin{figure}[th!]
\begin{center}
\includegraphics[width=7cm]{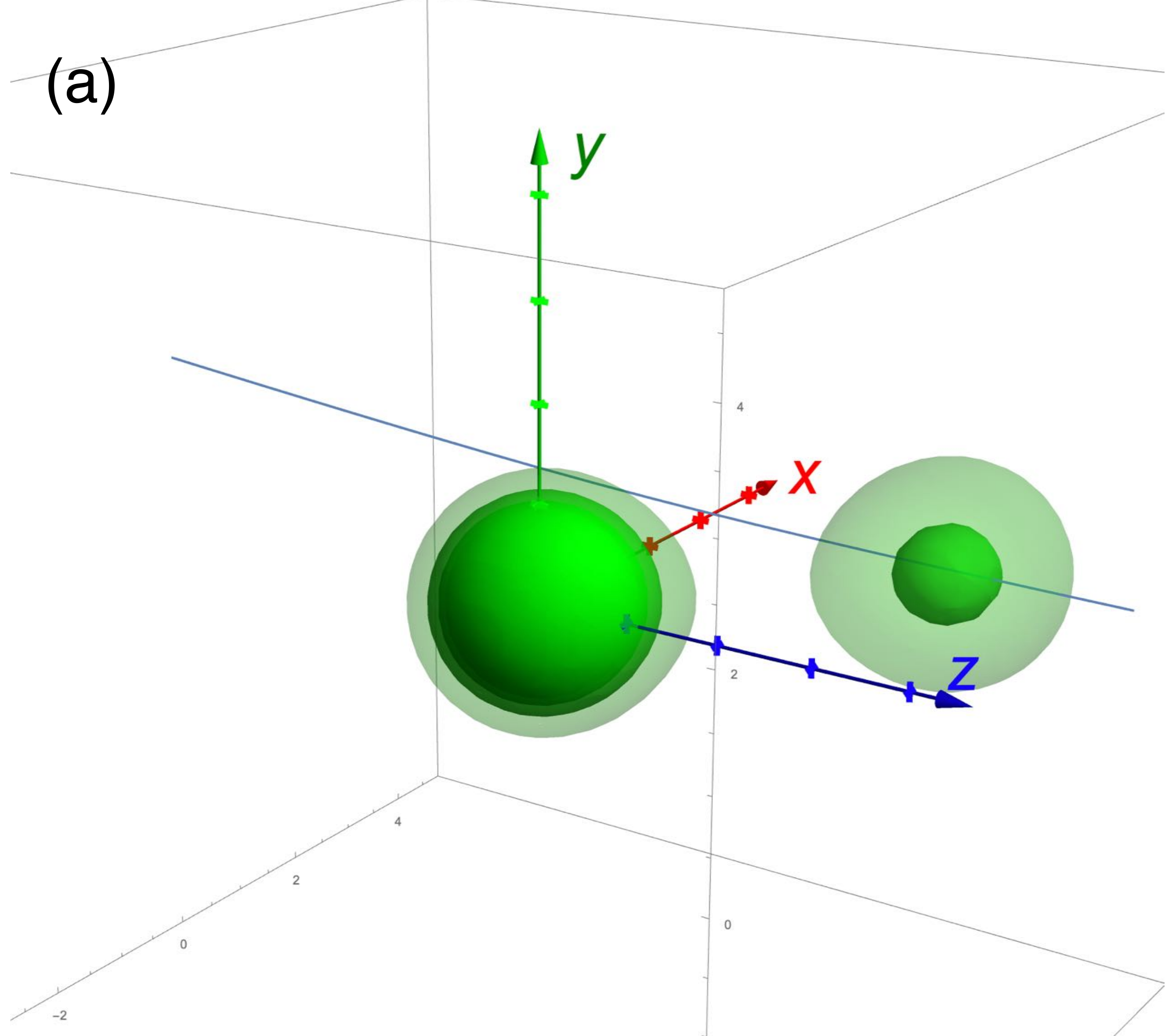} ~ \includegraphics[width=7cm]{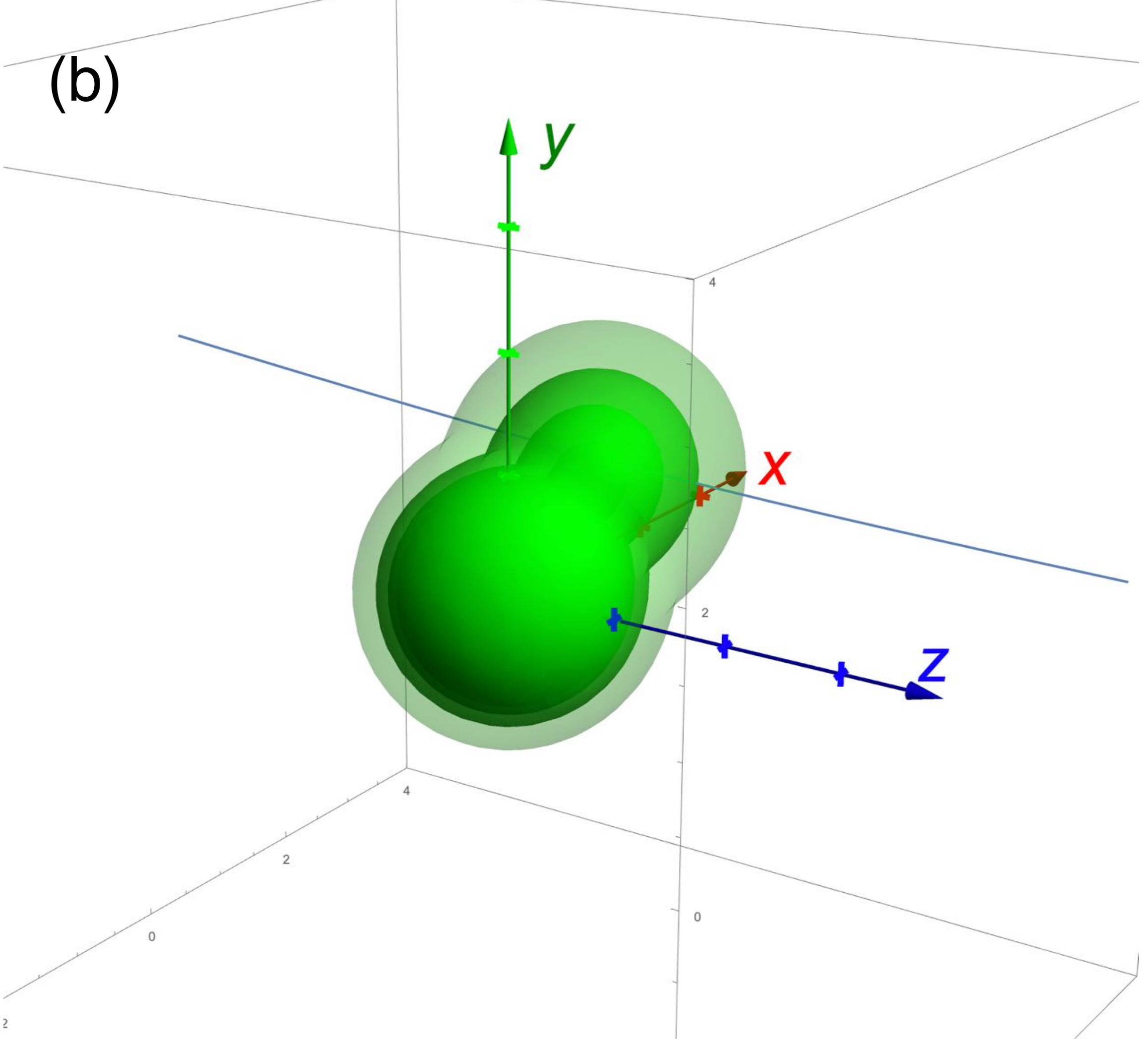} \\  
\includegraphics[width=7cm]{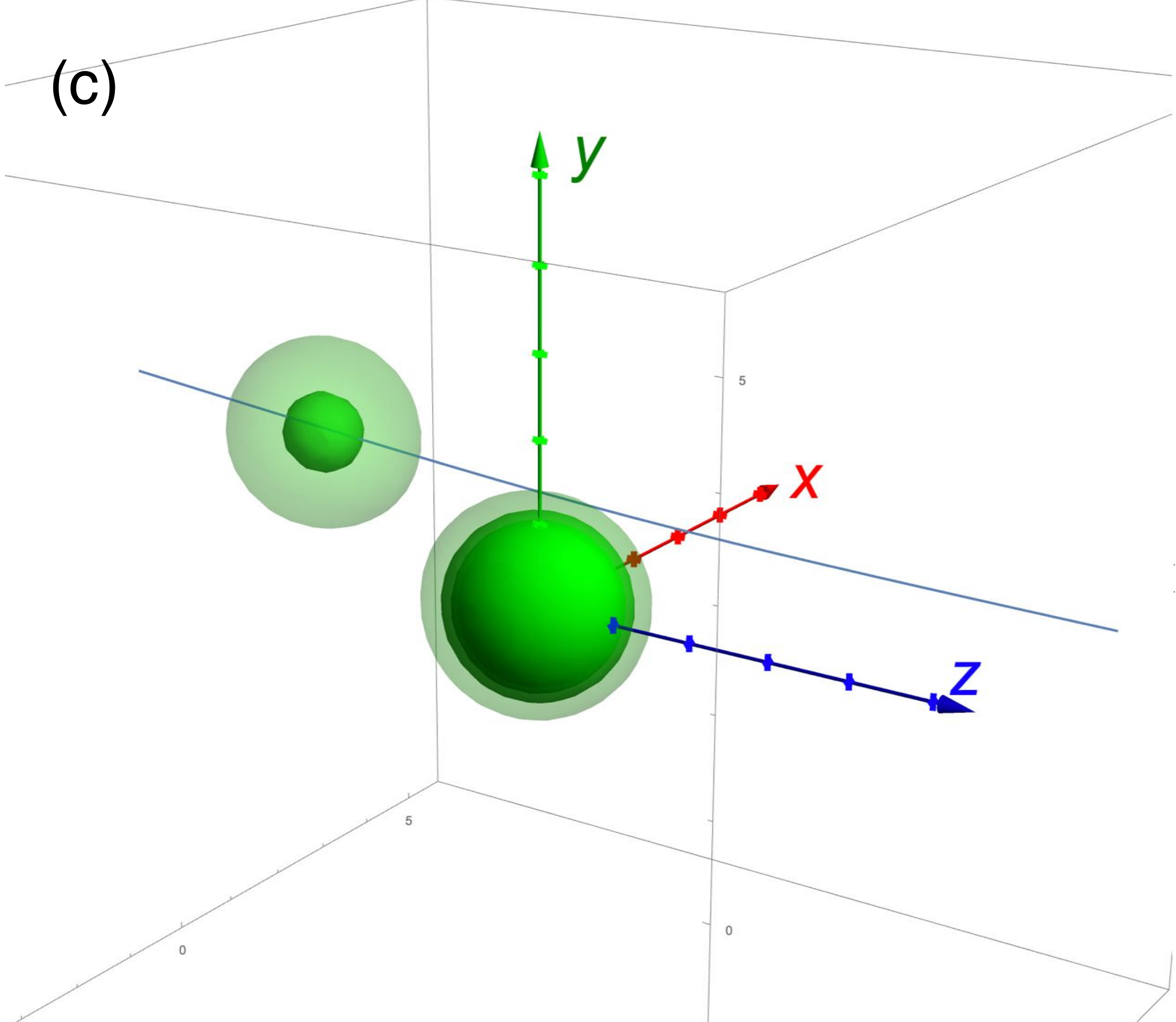} ~ \includegraphics[width=7cm]{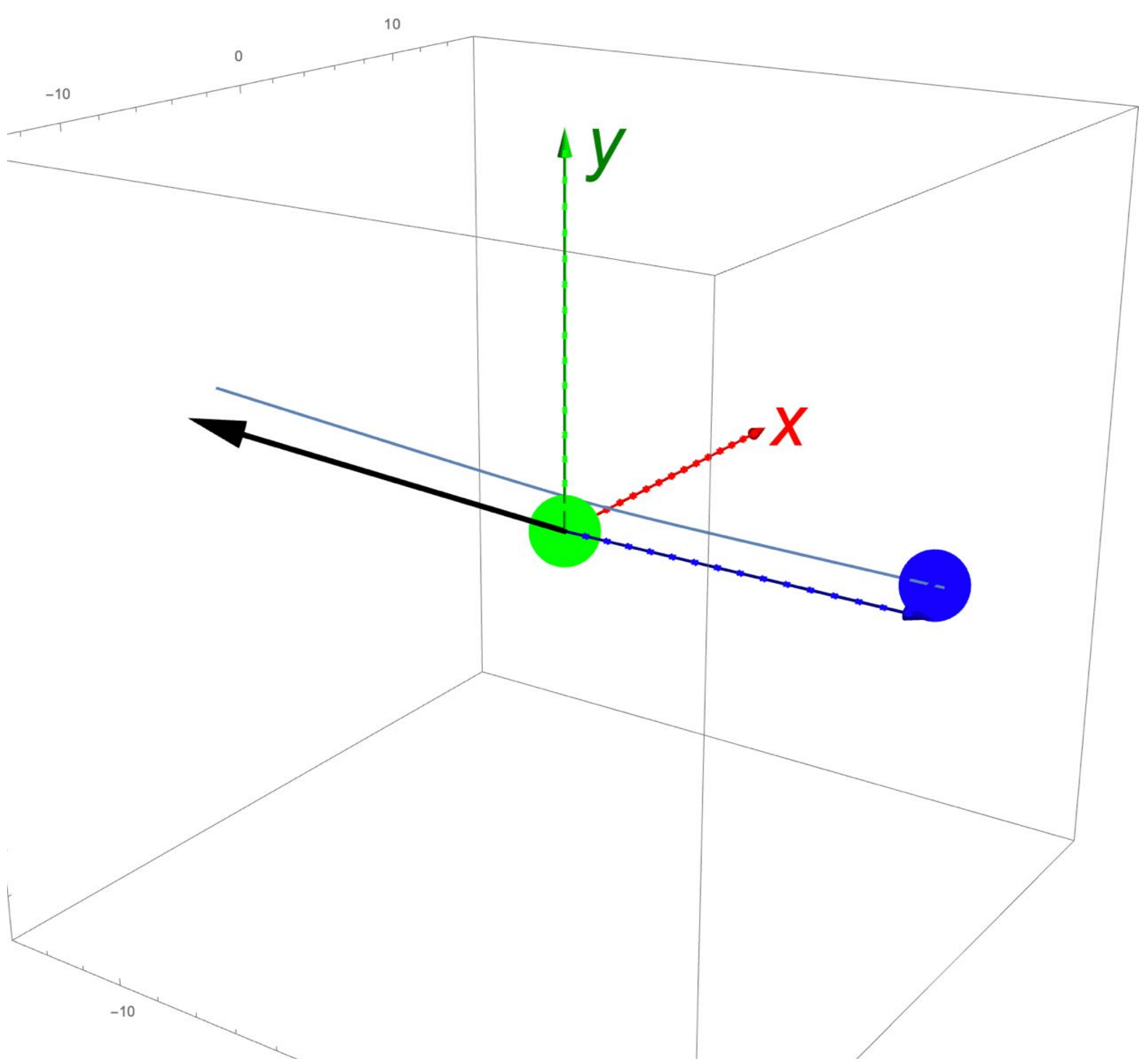}
\end{center}
\caption{Three frames from a scattering animation and a plot of the full trajectory with the with the outward direction $\hat{n}_{\rm out}'$ indicated.  The configuration has $k_{\rm t} = 2$, $\qq = 1$, and $C = 2$.  The initial position and momentum are $\vec{\RR}_0 = (3,0,15)$ and $\vec{\pp}_0 = (0,0,-4)$.  The simulation runs for 8 time units and the three frames shown are at $\tau = 3,4,5$ for (a), (b), (c) respectively.  The outward direction has unit vector $\hat{n}_{\rm out}' = (-0.52,0.30,-0.80)$ and is indicated by the black arrow in the last panel.  The outbound trajectory becomes parallel to this direction.}
\label{fig:sim4frames}
\end{figure}

The ancillary material includes an animation of a scattering trajectory.  In Figure \ref{fig:sim4frames} we show three frames from that animation and illustrate how the outward trajectory becomes parallel to the direction specified by \eqref{nout}.

\section{Conclusions}\label{sec:Conclusions}

In this work we have analyzed, both numerically and analytically, the interactions of a BPS monopole with an arbitrary number of 't Hooft defects.  Our motivations were to
\begin{enumerate}
\item broaden our understanding of classical soliton dynamics in the presence of defect singularities, and
\item illustrate with a new class of examples the emergence of particle dynamics from field theory through the collective coordinate paradigm for solitons.
\end{enumerate} 

Our main numerical results consist of simulations built on two key inputs.  First, the monopole and defect positions are represented in three-dimensional plots based on the energy density of the fields, determined analytically from the Blair--Cherkis--Durcan solutions.  We plot several level sets of the energy density with varying opacity, illustrating finiteness of the  density in the core of the smooth monopole and divergences in the cores of the defects.  Second, motion of the smooth monopole is generated by numerical integration of the equations of motion determined from the collective coordinate reduction to monopole moduli space.  The Mathematica code developed to produce the animations, as well as several example movies have been included with this submission as ancillary files.

In Section \ref{sec:Analytic} we explored the case of a single defect analytically, building on the work of \cite{Lee:2000rp,Jante:2015xra}.  We determined the period of an elliptical orbit, and a provided a new and elementary analysis of the scattering problem.

It would be interesting to extend the numerical techniques of this paper to the case of multi-monopole interactions in models based on a higher rank gauge group.  In such theories, monopoles come in different types because there are distinct types of magnetic charges they can carry---as many as the rank of the gauge group \cite{Weinberg:1979zt}.  Furthermore, the classical solutions and moduli space geometry for multi-monopole configurations with constituents of distinct type are much more tractable than for multi-monopole configurations carrying only one type of magnetic charge.  See, \eg~\cite{Lee:1996kz,Weinberg:1998hn}.  Simulating the dynamics for generic initial conditions should be possible, and this is an area in the field of magnetic monopoles that has not yet been explored.

\section*{Acknowledgments}

We thank Sergey Cherkis and the anonymous referee for helpful comments on the manuscript, and we thank Chris Halcrow and Ilarion Melnikov for illuminating discussions.  The work of GEL and KEW was supported in part by 2020 Erickson Discovery Grants for Undergraduate Research though the Pennsylvania State University.  The authors also thank the Division of Faculty Affairs at Penn State Fayette for support in early stages of the project.






\providecommand{\href}[2]{#2}\begingroup\raggedright\endgroup

\end{document}